\documentclass[twocolumn,showpacs,aps,epsfig,nofootinbib]{revtex4}

\usepackage[pdftex]{graphicx}
\usepackage{epstopdf}
\usepackage{latexsym}
\usepackage{amssymb}
\usepackage{color}
\newcommand{\nequation}{\setcounter{equation}{0}}
\renewcommand{\theequation}{\mbox{\arabic{section}.\arabic{equation}}}
\newcommand{\proofbegin}{\noindent{\it Proof.\,\,}}
\newcommand{\proofend}{\hfill$\Box$\bigskip}

\newtheorem{theorem}{Theorem}[section]
\newtheorem{proposition}[theorem]{Proposition}

\usepackage[center]{subfigure}

\begin{document}

 \newcommand{\bq}{\begin{equation}}
 \newcommand{\eq}{\end{equation}}
 \newcommand{\bqn}{\begin{eqnarray}}
 \newcommand{\eqn}{\end{eqnarray}}
 \newcommand{\nb}{\nonumber}
 \newcommand{\lb}{\label}
\newcommand{\PRL}{Phys. Rev. Lett.}
\newcommand{\PL}{Phys. Lett.}
\newcommand{\PR}{Phys. Rev.}
\newcommand{\CQG}{Class. Quantum Grav.}

\title{Gravitational collapse in Ho\v{r}ava-Lifshitz theory}

\author{Jared Greenwald ${}^{a}$}
\email{Jared_Greenwald@baylor.edu}

\author{Jonatan Lenells ${}^{b}$}
\email{Jonatan_Lenells@baylor.edu}

\author{V. H. Satheeshkumar ${}^{a}$}
\email{VH_Satheeshkumar@baylor.edu}

\author{Anzhong Wang ${}^{a, c}$}
\email{Anzhong_Wang@baylor.edu}

\affiliation{${}^{a}$ GCAP-CASPER, Physics Department, Baylor
University, Waco, TX 76798-7316, USA \\
${}^{b}$ Department of Mathematics, Baylor
University, Waco, TX 76798-7328, USA\\
${}^{c}$ Institute  for Advanced Physics $\&$ Mathematics,   Zhejiang University of
Technology, Hangzhou 310032,  China}

\date{\today}
\begin{abstract}
We study gravitational collapse of  a spherical fluid in nonrelativistic general covariant theory of the Ho\v{r}ava-Lifshitz 
gravity  with the projectability condition and an arbitrary coupling constant $\lambda$, where $|\lambda - 1|$ characterizes 
the deviation of the theory from general relativity in the infrared limit. The junction conditions  across the surface of  a 
collapsing star are derived under the (minimal) assumption that the junctions be mathematically meaningful in terms of 
distribution theory.  When the collapsing star is made of a homogeneous and isotropic perfect fluid,  and the external 
region is described by a stationary spacetime, the problem reduces to the matching of six independent conditions. If
the perfect fluid is pressureless (a dust fluid), it is found that the matching is also possible. In particular, in the case 
$\lambda  = 1$,  the external spacetime is described by the Schwarzschild (anti-) de Sitter solution written in 
Painlev\'e-Gullstrand coordinates. In the case $\lambda  \not= 1$,  the external spacetime is static but not asymptotically 
flat.  Our treatment can be easily generalized to other versions of Ho\v{r}ava-Lifshitz gravity or, more generally, to any 
theory of  higher-order derivative gravity.

\end{abstract}


\maketitle

\section{Introduction}
\nequation

The study of gravitational collapse provides useful insights into the {final} fate of a massive star \cite{Joshi}. Within the framework of general relativity, the dynamical collapse of a 
homogeneous spherical dust cloud under its own gravity was first considered by Datt \cite{Datt} and Oppenheimer and Snyder \cite{OppSnyder}. It was shown that it always leads to the formation of singularities. However, in a theory 
of quantum gravity,  it is expected that the formation of singularities in a gravitational collapse is prevented by short-distance quantum effects.

 In this paper,  we study this phenomenon (classically)  in the context of  the  Ho\v{r}ava theory of  gravity   \cite{Horava}. Since Ho\v{r}ava's theory is motivated by 
the Lifshitz theory in solid state physics \cite{Lifshitz}, it is often referred to as Ho\v{r}ava-Lifshitz (HL) theory.
One of the essential ingredients of the theory is the inclusion of higher-dimensional spatial derivative operators which dominate in the ultraviolet, making the theory power-counting renormalizable. The exclusion of higher-dimensional time
derivative operators, on the other hand, guarantees that the theory is unitary (the problem of non-unitarity has plagued the quantization of 
gravity for a long time \cite{Stelle}). However, this asymmetrical treatment of the space and time variables inevitably leads to the breaking of Lorentz symmetry. Although such a breaking is much less restricted by experiments in the gravitational sector than it is in the matter sector  \cite{LZbreaking,Pola}, the question of how to prevent the propagation of the Lorentz violations into the Standard Model of particle physics remains challenging \cite{PS}.

The breaking of Lorentz symmetry in the  ultraviolet  manifests itself in strongly anisotropic scalings of space and time, 
\bq
\lb{1.1}
{\bf x} \rightarrow \ell {\bf x}, \;\;\;  t \rightarrow \ell^{z} t.
\eq
In $(3+1)$-dimensional spacetimes, HL theory is power-counting 
renormalizable provided that  $z \ge 3$ \cite{Horava,Visser}.   In this paper, we will assume that $z =3$. At low energies, the theory is expected to flow 
to $z = 1$. In this limit the Lorentz invariance is ``accidentally restored." 

The anisotropy between time and space mentioned above is conveniently expressed in terms of the Arnowitt-Deser-Misner (ADM) decomposition \cite{ADM},
$
N, \; N^{i}, \; g_{ij},  \; (i, \; j = 1, 2, 3),
$
which are, respectively, the lapse function, shift vector, and the three-dimensional metric defined on the leaves of constant time.   
The requirement that the foliation defined by these leaves be preserved by any gauge symmetry implies that the theory is covariant only under the action of the group Diff($M, \; {\cal{F}}$) of foliation-preserving diffeomorphisms,  
\bq
\lb{1.4}
\delta{t} =  - f(t),\; \;\; \delta{x}^{i}  =   - \zeta^{i}(t, {\bf x}).
\eq
As a consequence, an additional degree of freedom appears
in the gravitational sector -- the spin-0 graviton. In order to be consistent with observations, this degree of freedom needs to decouple in the infrared (IR). Whether this decoupling takes place or not is still an open question \cite{reviews}. Let us point out that the spin-0 mode is unstable in the Minkowski background in the original incarnation of HL theory \cite{Horava}. If the projectability condition
\bq
\lb{1.6}
N = N(t)
\eq
remains imposed, this instability persists in the generalization of HL theory in which additional higher-order operators are included \cite{SVW, WM} 
(although in this case the de Sitter spacetime is stable \cite{HWW}). Another potential complication of HL theory is that the theory becomes strongly coupled when energy is very low \cite{BPS}. However, as long as the theory is consistent with observations when the nonlinear effects are taken into account, this is not necessarily a problem (at least not classically). A careful analysis shows that the theory is consistent with observations in the vacuum spherically symmetry static  case \cite{Mukreview} and in the cosmological setting \cite{WWa,IM,GMW}. 

One way to overcome the above problems is to introduce an extra local $U(1)$ symmetry, so that the total symmetry of the theory is enlarged to  \cite{HMT}
\bq
\lb{symmetry}
 U(1) \ltimes {\mbox{Diff}}(M, \; {\cal{F}}).
\eq 
This is achieved by introducing a gauge field $A$ and a Newtonian prepotential $\varphi$.
One consequence of the $U(1)$ symmetry is that the spin-0 gravitons are eliminated 
\cite{HMT,WWb}. As a result, all problems related to them, such as  instability, strong coupling, and different propagation speeds in the gravitational sector,
 are resolved. The $U(1)$ symmetry was initially introduced in the case of $\lambda = 1$, but the formalism was soon extended to the case of any $\lambda$ \cite{daSilva,HW,LWWZ}.
 In the presence of a $U(1)$ symmetry, the consistency of HL theory with solar system tests and cosmology 
was systematically studied in \cite{GSW,AP, cosmo}. In particular, it was shown in \cite{LMW} that in order for the theory to be consistent with solar system 
tests, the gauge field $A$ and the Newtonian prepotential $\varphi$ must be part of the metric in the IR limit (this ensures that the line element 
$ds^2$ is a scalar not only under $ {\mbox{Diff}}(M, \; {\cal{F}})$ but also under the local U(1) symmetry).

Another possibility is to give up the projectability condition (\ref{1.6}). This opens up for new operators to be included in the action, in particular, operators involving $a_i \equiv N_{,i}/N$ \cite{BPS}. In this way, all the problems mentioned above can be avoided by properly choosing the coupling constants.  However, since this leads to a theory with more than 70 independent coupling constants \cite{KP}, it makes the theory's predictive power  questionable, although only five coupling constants are relevant in the infrared.

A non-trivial generalization of the enlarged symmetry (\ref{symmetry}) to the nonprojectable case $N = N(t, x)$ was recently presented in \cite{ZWWS,ZSWW}. It was shown that, as in general relativity, the only degree of freedom of the model in the gravitational sector is the spin-2 massless graviton. Moreover, thanks to the elimination of the spin-0 gravitons, the physically viable range for the coupling constants is considerably enlarged, in comparison  with the healthy extension  \cite{BPS}, where the extra U(1) symmetry is absent.    Furthermore, the number of independent coupling constants is dramatically reduced 
from more than 70 to 15.  The consistency of the model with cosmology was recently established in \cite{ZSWW,ZHW,WWZZ}. In the case with spherical symmetry, the model was shown to be consistent with solar system tests  \cite{LW}. In contrast to the projectable case, the consistency can be achieved without taking the gauge field $A$ and  Newtonian prepotential $\varphi$ to be part of the metric. Finally, the duality between this version of HL theory and a non-relativistic quantum field theory was analyzed in \cite{Karch},  and its embedding  in string theory were constructed in \cite{Karchb} (For other examples, see for example, \cite{JKLCY}).

In this paper, we study gravitational collapse of a spherical star with a finite radius in the HL theory with the projectability condition,  an arbitrary coupling constant $\lambda$, and the extra U(1) symmetry \cite{HMT,WWb,daSilva,HW}. In general relativity, there are two common approaches for such studies. One approach relies on Israel's junction conditions \cite{Israel}, which are essentially obtained by using the Gauss and Codazzi equations. An advantage of this method is that it can be applied to the case where the coordinate systems inside and outside a collapsing body are different \footnote{Although Israel's method was initially developed only for non-null hypersurfaces, it was later generalized to the null hypersurface case \cite{BI}. For a recent review of this method, we refer to \cite{WS} and references therein.}.
The other approach is originally due to Taub \cite{Taub} and relies on distribution theory. In this approach, although the coordinate systems inside and outside the collapsing stars 
are taken to be the same, the null-hypersurface case can be easily included. Taub's approach was widely used to study colliding gravitational waves and other related issues
in general relativity \cite{Wang90s}. 

In this paper, we follow Taub's approach, as it turns out to be more convenient when dealing with higher-order derivatives. Moreover, in contrast to the case of general relativity, the foliation structure of  the HL theory implies that the coordinate systems inside and outside of the collapsing star are unique. Thus, {also} from a technical point of view, Taub's method seems a natural choice for the study of a collapsing star with a finite radius in the HL theory. 
 
	The paper is organized as follows: In Sec. II, we give a brief introduction to the HL theory with the projectability condition, an arbitrary coupling constant $\lambda$, and an extra U(1) symmetry. In Sec. III, we write down the field equations relevant for a spherical spacetime filled with a fluid. In Sec. IV, we generalize these equations to include the case where an infinitesimal thin matter shell appears on the surface of a collapsing star, and give explicitly all the necessary junction conditions. This generalization is carried out  under the {only} assumption that the junctions should be mathematically meaningful in terms of generalized functions; {therefore, in this sense the generalization is the most general}. In Sec. V, we apply the junction conditions to the case where the collapsing star is made of a homogeneous and isotropic perfect fluid, while the external region is described by a stationary  spacetime. 
When the perfect fluid is pressureless (a dust fluid), we find that matching is possible for any choice of    $\lambda$, but with different external spacetimes.
In particular, when $\lambda = 1$,   the external spacetime
is described by the Schwarzschild (anti-) de Sitter solution written in Painlev\'e-Gullstrand coordinates \cite{GP}. 
In Sec. VI, we present our main results and conclusions. Two appendices are also included. In Appendix A, some relevant functions are given for the spherical case considered here, while in Appendix B,
proof of  Eqs.(\ref{Fpm0})-(\ref{Fdeltaeqs}) is provided. 

We would like to  emphasize that our approach can
be easily generalized to other versions of HL gravity or, more generally, to any model of a higher-order derivative gravity theory.

\section{general covariant HL theory}
\nequation

In this section, we give a brief introduction to HL theory with the projectability condition
(\ref{1.6}),  an arbitrary coupling constant $\lambda$ and the enlarged symmetry (\ref{symmetry}).  For details, we refer
readers to  \cite{HW}.   The fundamental variables are ($N, \; N^i, \; g_{ij}, \; A, \; \varphi$), which transform as
\bqn
\delta{N} &=& \zeta^{k}\nabla_{k}N + \dot{N}f + N\dot{f},\nb\\
\delta{N}_{i} &=& N_{k}\nabla_{i}\zeta^{k} + \zeta^{k}\nabla_{k}N_{i}  + g_{ik}\dot{\zeta}^{k}
+ \dot{N}_{i}f + N_{i}\dot{f}, \nb\\
\delta{g}_{ij} &=& \nabla_{i}\zeta_{j} + \nabla_{j}\zeta_{i} + f\dot{g}_{ij},\nb\\
\delta{A} &=& \zeta^{i}\partial_{i}A + \dot{f}A  + f\dot{A},\nb\\
\delta \varphi &=&  f \dot{\varphi} + \zeta^{i}\partial_{i}\varphi,
\eqn
under  Diff($M, \; {\cal{F}}$), and as
\bqn
\lb{BB}
\delta_{\alpha}A &=&\dot{\alpha} - N^{i}\nabla_{i}\alpha,\;\;\;
\delta_{\alpha}\varphi = - \alpha,\nb\\ 
\delta_{\alpha}N_{i} &=& N\nabla_{i}\alpha,\;\;\;
\delta_{\alpha}g_{ij} = 0 = \delta_{\alpha}{N},
\eqn
under the local U(1) symmetry, where $\alpha$ is   the generator of the U(1) symmetry.
The total action is given by
 \bqn \lb{2.4}
S &=& \zeta^2\int dt d^{3}x N \sqrt{g} \Big({\cal{L}}_{K} -
{\cal{L}}_{{V}} +  {\cal{L}}_{{\varphi}} +  {\cal{L}}_{{A}} +  {\cal{L}}_{{\lambda}} \nb\\
& & ~~~~~~~~~~~~~~~~~~~~~~ \left. + {\zeta^{-2}} {\cal{L}}_{M} \right),
 \eqn
where $g={\rm det}\,g_{ij}$, and
 \bqn \lb{2.5}
{\cal{L}}_{K} &=& K_{ij}K^{ij} -   \lambda K^{2},\nb\\
{\cal{L}}_{\varphi} &=&\varphi {\cal{G}}^{ij} \Big(2K_{ij} + \nabla_{i}\nabla_{j}\varphi\Big),\nb\\
{\cal{L}}_{A} &=&\frac{A}{N}\Big(2\Lambda_{g} - R\Big),\nb\\
{\cal{L}}_{\lambda} &=& \big(1-\lambda\big)\Big[\big(\nabla^{2}\varphi\big)^{2} + 2 K \nabla^{2}\varphi\Big].
 \eqn
Here   the coupling constant $\Lambda_{g}$, which acts like a three-dimensional cosmological
constant, has the dimension of (length)$^{-2}$. The 
Ricci and Riemann terms all refer to the three-metric $g_{ij}$.
 $K_{ij}$ is the extrinsic curvature, and ${\cal{G}}_{ij}$ is the 3-dimensional ``generalized"
Einstein tensor defined  by
 \bqn \lb{2.6}
K_{ij} &=& \frac{1}{2N}\left(- \dot{g}_{ij} + \nabla_{i}N_{j} +
\nabla_{j}N_{i}\right),\nb\\
{\cal{G}}_{ij} &=& R_{ij} - \frac{1}{2}g_{ij}R + \Lambda_{g} g_{ij}.
 \eqn
${\cal{L}}_{M}$ is the
matter Lagrangian density and  
${\cal{L}}_{{V}}$ denotes the potential part of the action given by
 \bqn \lb{2.5a} 
{\cal{L}}_{{V}} &=& \zeta^{2}g_{0}  + g_{1} R + \frac{1}{\zeta^{2}}
\left(g_{2}R^{2} +  g_{3}  R_{ij}R^{ij}\right)\nb\\
& & + \frac{1}{\zeta^{4}} \left(g_{4}R^{3} +  g_{5}  R\;
R_{ij}R^{ij}
+   g_{6}  R^{i}_{j} R^{j}_{k} R^{k}_{i} \right)\nb\\
& & + \frac{1}{\zeta^{4}} \left[g_{7}R\nabla^{2}R +  g_{8}
\left(\nabla_{i}R_{jk}\right)
\left(\nabla^{i}R^{jk}\right)\right],  ~~~~
 \eqn 
which preserves the parity,  where the coupling  constants $ g_{s}\, (s=0, 1, 2,\dots 8)$  are all dimensionless. The relativistic limit in the IR
 requires that
 \bq
 \lb{2.5b}
 g_{1} = -1,\;\;\; \zeta^2 = \frac{1}{16\pi G},
 \eq
where $G$ denotes the Newtonian constant.

Variation of the total action (\ref{2.4}) with respect to the lapse function $N(t)$  yields the
Hamiltonian constraint
 \bqn \lb{eq1}
& & \int{ d^{3}x\sqrt{g}\left[{\cal{L}}_{K} + {\cal{L}}_{{V}} - \varphi {\cal{G}}^{ij}\nabla_{i}\nabla_{j}\varphi 
- \big(1-\lambda\big)\big(\nabla^{2}\varphi\big)^{2}\right]}\nb\\
& & ~~~~~~~~~~~~~~~~~~~~~~~~~~~~~
= 8\pi G \int d^{3}x {\sqrt{g}\, J^{t}},
 \eqn
where
 \bq \lb{eq1a}
J^{t} = 2 \frac{\delta\left(N{\cal{L}}_{M}\right)}{\delta N}.
 \eq
 
Variation of the action with respect to the shift $N^{i}$ yields the
super-momentum constraint
 \bqn \lb{eq2}
& & \nabla^{j}\Big[\pi_{ij} - \varphi  {\cal{G}}_{ij} - \big(1-\lambda\big)g_{ij}\nabla^{2}\varphi \Big] = 8\pi G J_{i}, ~~~
 \eqn
where the super-momentum $\pi^{ij} $ and matter current $J^{i}$
are defined as
 \bq
 \lb{eq2a}
\pi^{ij} \equiv
 - K^{ij} + \lambda K g^{ij},\;\;
J^{i} \equiv - N\frac{\delta{\cal{L}}_{M}}{\delta N_{i}}.
 \eq
Similarly, variations of the action with respect to $\varphi$ and $A$ yield, respectively, 
\bqn
\lb{eq4a}
& & {\cal{G}}^{ij} \Big(K_{ij} + \nabla_{i}\nabla_{j}\varphi\Big) + \big(1-\lambda\big)\nabla^{2}\Big(K + \nabla^{2}\varphi\Big) \nb\\
& & ~~~~~~~~~~~~~~~~~~~~~~~~~ = 8\pi G J_{\varphi}, \\
\lb{eq4b}
& & R - 2\Lambda_{g} =    8\pi G J_{A},
\eqn
where
\bq
\lb{eq5}
J_{\varphi} \equiv - \frac{\delta{\cal{L}}_{M}}{\delta\varphi},\;\;\;
J_{A} \equiv 2 \frac{\delta\left(N{\cal{L}}_{M}\right)}{\delta{A}}.
\eq
On the other hand, variation with respect to $g_{ij}$ leads to the
dynamical equations
 \bqn \lb{eq3}
&&
\frac{1}{N\sqrt{g}}\Bigg\{\sqrt{g}\Big[\pi^{ij} - \varphi {\cal{G}}^{ij} - \big(1-\lambda\big) g^{ij} \nabla^{2}\varphi\Big]\Bigg\}_{,t} 
\nb\\
& &~~~ = -2\left(K^{2}\right)^{ij}+2\lambda K K^{ij} \nb\\
& &  ~~~~~ + \frac{1}{N}\nabla_{k}\left[N^k \pi^{ij}-2\pi^{k(i}N^{j)}\right]\nb\\
& &  ~~~~~ - 2\big(1-\lambda\big) \Big[\big(K + \nabla^{2}\varphi\big)\nabla^{i}\nabla^{j}\varphi + K^{ij}\nabla^{2}\varphi\Big]\nb\\
& & ~~~~~ + \big(1-\lambda\big) \Big[2\nabla^{(i}F^{j)}_{\varphi} - g^{ij}\nabla_{k}F^{k}_{\varphi}\Big]\nb\\
& & ~~~~~ +  \frac{1}{2} \Big({\cal{L}}_{K} + {\cal{L}}_{\varphi} + {\cal{L}}_{A} + {\cal{L}}_{\lambda}\Big) g^{ij} \nb\\
& &  ~~~~~    + F^{ij} + F_{\varphi}^{ij} +  F_{A}^{ij} + 8\pi G \tau^{ij},
 \eqn
where $\left(K^{2}\right)^{ij} \equiv K^{il}K_{l}^{j},\; f_{(ij)}
\equiv \left(f_{ij} + f_{ji}\right)/2$, and
 \bqn
\lb{eq3a} 
F_{A}^{ij} &=& \frac{1}{N}\left[AR^{ij} - \Big(\nabla^{i}\nabla^{j} - g^{ij}\nabla^{2}\Big)A\right],\nb\\ 
F_{\varphi}^{ij} &=&  \sum^{3}_{n=1}{F_{(\varphi, n)}^{ij}},\nb\\
F^{ij} &\equiv& \frac{1}{\sqrt{g}}\frac{\delta\left(-\sqrt{g} {\cal{L}}_{V}\right)}{\delta{g}_{ij}}
 = \sum^{8}_{s=0}{g_{s} \zeta^{n_{s}} \left(F_{s}\right)^{ij} }, \qquad
 \eqn
with 
$n_{s} =(2, 0, -2, -2, -4, -4, -4, -4,-4)$.  The  3-tensors $ \left(F_{s}\right)_{ij}$ and 
$F_{(\varphi, n)}^{ij}$ are given by Eqs.(2.21)-(2.23)
in \cite{WWb}, which, for the sake of the readers' convenience, are reproduced in  Eqs.(\ref{A.1}) and (\ref{A.2}) of this
paper. The stress 3-tensor $\tau^{ij}$ is defined as
 \bq \label{tau}
\tau^{ij} = {2\over \sqrt{g}}{\delta \left(\sqrt{g}
 {\cal{L}}_{M}\right)\over \delta{g}_{ij}}.
 \eq
 
The matter quantities $(J^{t}, \; J^{i},\; J_{\varphi},\; J_{A},\; \tau^{ij})$ satisfy the
conservation laws
 \bqn \lb{eq5a} & &
 \int d^{3}x \sqrt{g} { \left[ \dot{g}_{kl}\tau^{kl} -
 \frac{1}{\sqrt{g}}\left(\sqrt{g}J^{t}\right)_{, t}  
 +   \frac{2N_{k}}  {N\sqrt{g}}\left(\sqrt{g}J^{k}\right)_{,t}
  \right.  }   \nb\\
 & &  ~~~~~~~~~~~~~~ \left.   - 2\dot{\varphi}J_{\varphi} -  \frac{A} {N\sqrt{g}}\left(\sqrt{g}J_{A}\right)_{,t}
 \right] = 0,\\
\lb{eq5b} & & \nabla^{k}\tau_{ik} -
\frac{1}{N\sqrt{g}}\left(\sqrt{g}J_{i}\right)_{,t}  - \frac{J^{k}}{N}\left(\nabla_{k}N_{i}
- \nabla_{i}N_{k}\right)   \nb\\
& & \;\;\;\;\;\;\;\;\;\;\;- \frac{N_{i}}{N}\nabla_{k}J^{k} + J_{\varphi} \nabla_{i}\varphi - \frac{J_{A}}{2N} \nabla_{i}A
 = 0.
\eqn

In general relativity,  the four-dimensional energy-momentum tensor is defined as
\bq
\lb{EMT}
T^{\mu\nu} = \frac{1}{\sqrt{-g^{(4)}}} \frac{\delta\left(\sqrt{-g^{(4)}}{\cal{L}}_{M}\right)}{\delta g^{(4)}_{\mu\nu}},
\eq
where $\mu, \nu = 0, 1, 2, 3$, and 
\bq
\lb{4Dmetric}
g^{(4)}_{00} = -N^{2} + N^{i}N_{i}, \;\;\; g^{(4)}_{0i} = N_{i}, \;\;\; g^{(4)}_{ij} = g_{ij}. 
\eq
Introducing the normal vector $n_{\mu}$ to the hypersurface $ t = $ constant by
\bq
\lb{4Dnorm}
n_{\mu} = N\delta^{t}_{\mu}, \;\;\; n^{\mu} = \frac{1}{N} \left(- 1,   N^{i} \right),
\eq
one can decompose $T_{\mu\nu}$ as follows \cite{Ann}:
\bqn
\lb{4DEM}
\rho_{H} &\equiv& T_{\mu\nu} n^{\mu} n^{\nu},\;\;\;
s_{i}  \equiv -  T_{\mu\nu} h^{(4)\mu}_{i} n^{\nu},\nb\\ 
s_{ij}  &\equiv&  T_{\mu\nu} h^{(4)\mu}_{i} h^{(4)\nu}_{j},
\eqn
where $h^{(4)}_{\mu\nu}$ is the projection operator defined by $h^{(4)}_{\mu\nu} \equiv g^{(4)}_{\mu\nu}
+ n_{\mu}n_{\nu}$. In the relativistic  limit, one may make the following  identification: 
\bq
 \left(J^{t},\; J_{i}, \; \tau_{ij}\right) = \left(- 2\rho_{H},\; - s_{i},\; s_{ij}\right).
 \eq

\section{ Spherical Spacetimes Filled with a fluid}
\nequation

Spherically symmetric static spacetimes in the framework of the HL theory with U(1) symmetry   with or without 
 the projectabilty condition are studied systematically in  \cite{GSW,AP,GLLSW,BLW,LMW,LW,AK}.  In particular, the
 ADM variables  for   spherically symmetric spacetimes    with the projectability condition take the  forms
\bqn
\lb{3.1b}
&& N = 1, \;\;\; N^i = \delta^{i}_{r} e^{\mu(r, t) - \nu(r, t)},\nb\\
&& g_{ij}dx^idx^j =  e^{2\nu(r, t)} dr^2   + r^{2}d\Omega^2,
\eqn
in the spherical coordinates $x^{i} = (r, \theta, \phi)$, where $d\Omega^2 \equiv
d\theta^{2}  + \sin^{2}\theta\, d\phi^{2}$. The diagonal case $N^i = 0$ corresponds to $\mu(t, r) = -\infty$. {On the other hand, using}  
the $U(1)$ gauge freedom (\ref{BB}), without loss of generality, we set
\bq
\lb{gauge}
\varphi = 0,
\eq
which  uniquely fixes the gauge. Then, we  find that  
 \bqn
\lb{3.3a}
{\cal{L}}_{\varphi}  &=& 0 = {\cal{L}}_{\lambda},\;\;\; F^{ij}_{\varphi} = 0,\nb\\
K_{ij} &=& e^{\mu+\nu}\Big((\mu'-\dot{\nu} e^{-\mu+\nu}) \delta^{r}_{i}\delta^{r}_{j} + re^{-2\nu}\Omega_{ij}\Big),\nb\\
R_{ij} &=&  \frac{2\nu'}{r}\delta^{r}_{i}\delta^{r}_{j} + e^{-2\nu}\Big[r\nu' - \big(1-e^{2\nu}\big)\Big]\Omega_{ij},\nb\\
{\cal{L}}_{K} &=& (1 - \lambda) \Bigg[\dot{\nu}^2 - 2\dot{\nu} \mu' e^{\mu-\nu} + \left( {\mu'}^2 + \frac{2}{r^{2}} \right) e^{2(\mu-\nu)} \Bigg]\nb\\
&& + \lambda \left[ \frac{4}{r} \dot{\nu} e^{\mu-\nu} - \frac{2}{r^{2}} e^{2(\mu-\nu)}\left(2r\mu' + 1\right)\right]  \nb\\ 
{\cal{L}}_{A} &=&  \frac{2A}{r^2} \Big[e^{-2\nu}\left(1 - 2r \nu'\right) + \Lambda_{g} r^2 - 1\Big],\nb\\
{\cal{L}}_{V} &=& \sum_{s=0}^{3}{{\cal{L}}_{V}^{(s)}},
\eqn
where  a prime denotes the partial derivative with respect to $r$, 
  $\Omega_{ij} \equiv \delta^{\theta}_{i}\delta^{\theta}_{j}  + \sin^{2}\theta\delta^{\phi}_{i}\delta^{\phi}_{j}$,
  and ${\cal{L}}_{V}^{(s)}$'s are given by Eq.(A1) in \cite{GPW}. 
The Hamiltonian constraint (\ref{eq1}) reads
 \bq 
 \lb{3.3b}
\int{\left( {\cal{L}}_{K} + {\cal{L}}_{V}  - 8 \pi G J^{t} \right) e^{\nu} r^{2} dr}
= 0,
 \eq 
 while the momentum constraint (\ref{eq2}) yields
 \bqn
 \lb{3.3c}
  && (1-\lambda)\Big\{e^{\mu - \nu}\left[r^2 (\mu'' + \mu'^2 - \mu' \nu') + 2 (\mu' r - 1)\right] \nb\\
  && ~~~~~~~~~~~  - \dot{\nu}' r^2\Big\}  + 2 r \left(\lambda \nu' e^{\mu - \nu} - \dot{\nu} \right)  \nb\\
  && ~~~~~~~~~~~  =  - 8 \pi G r^2 e^{-\mu + \nu}  v,
 \eqn
 where 
  \bqn 
\nb
 J^{i} \equiv e^{-(\mu + \nu)}\big(v, 0, 0\big).
 \eqn
 It can also be shown that Eqs.(\ref{eq4a}) and (\ref{eq4b}) now read
 \bqn
 \lb{3.3e}
& &  \Big[e^{2\nu}\left(\Lambda_{g}r^2 -1\right) + 1\Big] \Big( e^{\mu + \nu}\mu' - e^{2 \nu} \dot{\nu} \Big)\nb\\
&& ~ -2\Big(\nu' - \Lambda_{g}re^{2\nu}\Big) e^{\mu + \nu} \nb\\
&& + (1-\lambda)\Big\{e^{2\nu}\left( -r^2 \dot{\nu}'' + r^2 \dot{\nu}' \nu' - 2 r \dot{\nu}' \right) \nb\\
&& + e^{\mu + \nu} \left[ r^2 ({\mu}''' + 3 \mu' {\mu}'' - {\mu}' \nu'' - 3 {\mu}'' \nu' + \mu'^3 - 3 \nu' \mu'^2 \right. \nb\\ 
&&\left. + 2 \mu' \nu'^2 ) + 2r(2 \mu'' -\nu'' + 2 (\nu' -\mu')^2 ) + 2 \nu' \right]\Big\}\nb\\ 
&&= 8\pi G r^{2} e^{4\nu} J_{\varphi}, ~~~~~~~~ \\
\lb{3.3f}
& & 2 r \nu'  - \Big[e^{2\nu}\left(\Lambda_{g}r^2 - 1\right) +  1\Big] =   4\pi G r^2 e^{2\nu}  J_{A}.
\eqn
 The dynamical equations (\ref{eq3}), on the other hand, yield
 \bqn
 \lb{3.3g}
 && (1-\lambda)r \left[ e^{\nu+\mu} \big( \dot{\mu} \mu' + \dot{\mu}' -\dot{\nu}' \big) - e^{2\nu} \big( \ddot{\nu} + \frac{1}{2}{\dot{\nu}}^{2}\big) \right. \nb \\  
 && \left. + e^{2\mu} \big( \mu'' + \frac{1}{2}\mu'^{2} - \mu' \nu' \big) \right] \nb \\ 
& & + \left[2 \big(\mu' + \lambda \nu'\big)+ (4 \lambda - 3)\frac{1}{r}\right]e^{2\mu}  -2e^{\nu+\mu} \big(\lambda \dot{\mu} + \dot{\nu} \big) \nb\\
 & & + \frac{1}{2} r e^{2\nu} {\cal{L}}_{A} = - r\Big(F_{rr}  + F^{A}_{rr} + 8\pi G e^{2\nu}p_{r}\Big), \\
 \lb{3.3h}
  & &  \left[\lambda r \Big( \mu'' - \mu' \nu' \Big) + (2\lambda - 1) \big(2 \mu' - \nu' \big) \right.\nb\\
  && \left. ~ + \frac{1}{2} (3\lambda + 1) r \mu'^2 \right]e^{2\mu}  + \frac{1}{2} r e^{2\nu} {\cal{L}}_{A}  \nb\\
  && ~ + \Big( \lambda  \ddot{\nu} + \frac{1}{2} (\lambda + 1)  \dot{\nu}^2 \Big) r e^{2\nu}  \nb\\
  & & ~ - \left[(2\lambda - 1) \dot{\mu} + r \mu' \big( \dot{\nu} + \lambda \dot{\mu} \big) + \lambda r \big( \dot{\nu}'+ \dot{\mu}' \big) \right] e^{\nu + \mu} \nb\\
 & &  = -\frac{e^{ 2\nu}}{r}\left(F_{\theta\theta} + F^{A}_{\theta\theta} + 8\pi G r^{2}p_{\theta}\right),
 \eqn
 where
 \bqn
 \tau_{ij} &=& e^{2\nu}p_{r}\delta^{r}_{i}\delta^{r}_{j} + r^{2}p_{\theta}\Omega_{ij},\nb\\ 
F^{A}_{ij} &=&\frac{ 2}{r}\big(A' + A\nu'\big) \delta^{r}_{i}\delta^{r}_{j}  +  e^{-2\nu}\Big[r^{2}\big(A'' - \nu'A'\big)\nb\\ 
& & ~~~ + r\big(A' + A\nu'\big)  - A\Big(1 - e^{2\nu}\Big)\Big]\Omega_{ij}, 
\eqn
 {and } { $F_{ij}$   is given by Eq.(A4) in Appendix A.}
We define a fluid with $p_{r} = p_{\theta}$
as a perfect fluid, which in general allows  energy flow along a radial direction, i.e.,  $v$ does not not necessarily vanish \cite{Santos85}. 
 
The energy conservation law (\ref{eq5a}) now reads
\bqn
\lb{3.3ja}
\int dr \,\, e^\nu r^2 \Big[ \dot{\rho}_H + \left( \rho_H + 4 p_r\right)\dot{\nu}  ~~~~~~~~~~~~~~~~~~~~ \nb \\ 
 \quad  + 4 \left( \dot{v} - v\dot{\mu} \right) 
 - 2\left( \dot{J}_A + \dot{\nu} J_A\right) \Big] = 0, \qquad
\eqn
while the momentum conservation (\ref{eq5b}) yields
\bqn
\lb{3.3jb}
&& v\mu' - \big(v' - p_{r}'\big) - \frac{2}{r}\big(v - p_{r} + p_{\theta}\big)  - \frac{1}{2}J_{A} A' \nb\\
 && ~~~~ - e^{\nu-\mu} \Big[\dot{v} + v \big(2\dot{\nu} - \dot{\mu}\big)\Big] = 0.
\eqn

 To relate the quantities $J^{t},\; J^{i}$ and $\tau_{ij}$   to the ones often used in general relativity, 
 in addition to the normal vector $n_{\mu}$ defined in Eq.(\ref{4Dnorm}), we also introduce  
the spacelike unit vectors $\chi_{\mu}, \; \theta_{\mu}$ and $\phi_{\mu}$ by
\bqn
\lb{emt1}
 n_{\mu} &=& \delta^{t}_{\mu}, \;\;\; n^{\mu} = - \delta_{t}^{\mu} + e^{\mu-\nu}\delta_{r}^{\mu},\nb\\
 \chi^{\mu} &=& e^{-\nu}\delta^{\mu}_{r} , \;\;\; \chi_{\mu} = e^{\mu}\delta^{t}_{\mu} + e^{\nu} \delta^{r}_{\mu},\nb\\
 \theta_{\mu} &=& r\delta^{\theta}_{\mu},\;\;\; \phi_{\mu} = r\sin\theta \delta^{\phi}_{\mu}.
\eqn
In terms of these four unit  vectors,    
the energy-momentum tensor for an anisotropic fluid can be written as
\bqn
\lb{emt2}
T_{\mu\nu} &=& \rho_{H}n_{\mu}n_{\nu} + q \big(n_{\mu} \chi_{\nu} + n_{\nu} \chi_{\mu} \big)\nb\\
& & + p_{r}\chi_{\mu} \chi_{\nu}  + p_{\theta}\big(\theta_{\mu}\theta_{\nu} + \phi_{\mu}\phi_{\nu}\big),
\eqn
where $\rho_{H}, \; q,\; p_{r}$ and $p_{\theta}$ denote, respectively, the energy density, heat flow 
along radial direction, radial, and tangential pressures, {as } measured by the observer with the four-velocity
$n_{\mu}$. 
This decomposition is consistent with 
 the quantities $J^{t}$ and $J^{i}$ defined by
 \bq \lb{rhoHv}
 \rho_{H} = -\frac{1}{2} J^{t},\qquad v= e^{\mu} q. 
\eq
  It should be noted that the definitions of the energy density $\rho_H$, 
 the radial pressure $p_r$ and the heat flow $q$ are different from the ones defined in a comoving frame 
 in general relativity. We refer readers to Appendix B of \cite{GPW} for details.

\section{ Junction Condition Across the Surface of a Collapsing Sphere }
\nequation

\begin{figure}[tbp]
\centering
\includegraphics[width=6cm]{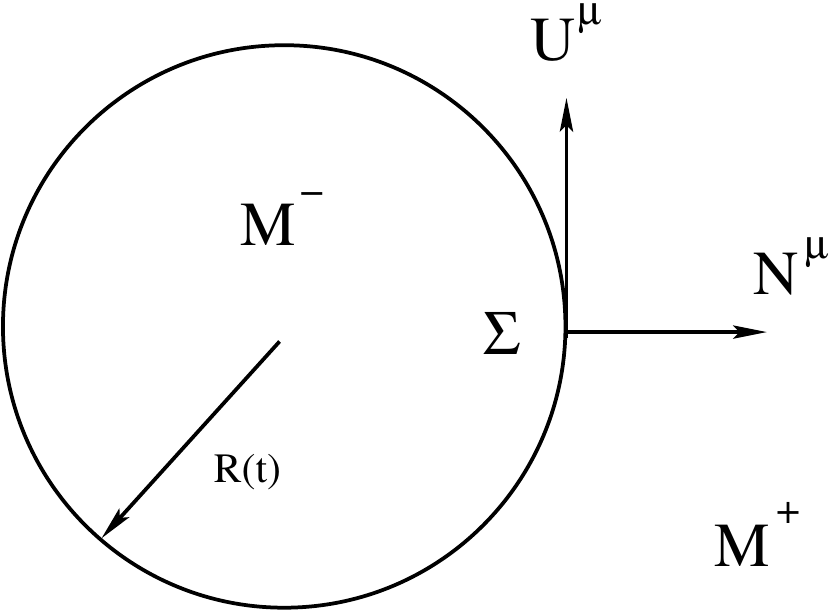} 
\caption{The spacetime is divided into two regions, the internal $M^{-}$ and external $M^{+}$, where 
$M^{-} = \left\{x^{\mu}: r < {\cal{R}} (t)\right\}$, and $M^{+} = \left\{x^{\mu}: r > {\cal{R}}(t)\right\}$. The surface $r = {\cal{R}}(t)$ is denoted by $\Sigma$.}
\label{fig1}
\end{figure}

The surface $\Sigma$ of a spherically symmetric collapsing star naturally divides the spacetime $M$ into 
two regions, the internal and the external regions, denoted by
$M^-$ and $M^+$ respectively, as shown schematically in Fig. \ref{fig1}. 
The surface $\Sigma = \partial M^- = -\partial M^+$ is described by
\bq
\lb{5.1}
\Phi(t, r) = 0,
\eq
where $\Phi(t,r) \equiv r - {\cal{R}}(t)$. The spherical symmetry implies that the ADM variables  on $M$ take the form (\ref{3.1b}).

\subsection{Preliminaries}
We assume that the normal vector $\nabla \Phi$ to the hypersurface $\Sigma$ with components
\bqn \label{gradPhicomponents}
&& \Phi_{,\lambda} = \delta_{\lambda}^{r} - \dot{\cal{R}} \delta^{t}_{\lambda}, 
 	\\ \nb
&& \Phi^{,\lambda} = e^{-2\nu}(1 - e^{2\mu} - \dot{\mathcal{R}} e^{\mu + \nu}) \delta^{\lambda}_{r} + (e^{\mu - \nu} + \dot{\mathcal{R}}) \delta_{t}^{\lambda},
\eqn
is everywhere spacelike, i.e.
\bq
\lb{5.3a}
  \Phi^{,\lambda}\Phi_{,\lambda} = e^{-2\nu}\bigl[1 - (e^\mu + e^\nu \dot{\mathcal{R}})^2\bigr] > 0.
\eq
This is the case if $\dot{\mathcal{R}}$ is small enough. We may then define the vector field $N = \nabla \Phi/\|\nabla \Phi\|_g$ in a neighborhood of $\Sigma$.\footnote{It must not be confused with the
lapse function, as in the present case it is set to one, as one can see from Eq.(\ref{3.1b}).}    $N$ has length one, i.e. $N_{\lambda}N^\lambda = 1$, and the restriction of $N$ to $\Sigma$ is the 
outward pointing unit normal vector field on $\Sigma$.

Let $H(\Phi)$ denote the Heaviside function defined by
\bq
\lb{5.3}
H(\Phi) = \cases{1, & $\Phi > 0$,\cr
\frac{1}{2}, & $\Phi = 0$,\cr
0, & $ \Phi < 0$,\cr}
\eq
and let $\delta(\Phi)$ denote the delta distribution with support on $\Sigma$.
By definition, $\delta(\Phi)$ acts on a smooth test function $\varphi \in C^\infty(M)$ of compact support by
\bq
(\delta(\Phi), \varphi) = \int_\Sigma \varphi d\Sigma,
\eq
where $d\Sigma = \iota_N \text{Vol}_g$ is the volume three-form induced by $g$ on $\Sigma$ and $\iota_N$ denotes interior multiplication by $N$. 
The derivatives $\delta^{(n)}(\Phi)$, $n \geq1$, of $\delta(\Phi)$ are defined in a standard way and the following relations are valid  \cite{GS1964}:
\bqn \nb
 \frac{\partial H(\Phi)}{\partial x^\lambda}
& = & \frac{\partial \Phi}{\partial x^\lambda} \delta(\Phi),
	\\ \nb
 \frac{\partial}{\partial x^\lambda} \delta^{(n)}(\Phi) & = & \frac{\partial \Phi}{\partial x^\lambda} \delta^{(n+1)}(\Phi), \qquad n = 0, 1,2, \dots,
 	\\ \label{deltarecursion}
\Phi \delta^{(n)}(\Phi) & =& -n \delta^{(n-1)}(\Phi), \qquad n = 1, 2, \dots.
\eqn

If $f$ is a function defined in a neighborhood of $\Sigma$, we define the distribution $f \delta^{(n)}(\Phi)$ by letting it act on a test function $\varphi$ by
\bq
(f \delta^{(n)}(\Phi), \varphi) = (\delta^{(n)}(\Phi), f \varphi).
\eq
The product $f\delta(\Phi)$ is well defined whenever $f$ is $C^0$ and it depends only on the restriction $f|_\Sigma$ of $f$ to $\Sigma$. 
More generally, the product $f\delta^{(n)}(\Phi)$ is well defined provided that $f$ is $C^n$ and it depends only on the values of $f$ and its 
partial derivatives of order $\leq n$ evaluated on $\Sigma$.

Let $F$ be a distribution on $M$ of the form
\bq\label{Fdef}
F = F^+ H(\Phi) + F^- [1-H(\Phi)] + \sum_{k = 0}^n F^{Im(k)} \delta^{(k)}(\Phi),
\eq
where the $F_n$'s are functions defined in a neighborhood of $\Sigma$ while $F^+$ and $F^-$ are sufficiently smooth functions defined on 
$M^+$ and $M^-$ respectively. We define the function $F^D$ on $M$ by
\bq
F^D = F^+ H(\Phi) + F^- [1-H(\Phi)],
\eq
and we define the jump $[F]^-$ of $F$ across $\Sigma$ by
\bq
[F]^-(x) = F^+(x) - F^-(x), \qquad x \in \Sigma.
\eq
We will also need the fact that the equation $F = 0$ is equivalent to the equations
\bqn\label{Fpm0}
F^\pm(x) = 0, \qquad x \in M^\pm,
\eqn
and
\bqn \nb
&& \sum_{k = 0}^j (-1)^k \frac{(n-k)!j!}{(j-k)!} \frac{\partial^{j-k}}{\partial \Phi^{j-k}} F^{Im(n-k)}\bigg|_{\Sigma} = 0,
	\\ \label{Fdeltaeqs}
&&\hspace{4cm} 0 \leq j \leq n,
\eqn
where $\frac{\partial}{\partial \Phi}$ acts on a function� $f$ by
\bqn\label{dfdPhi}
\frac{\partial f}{\partial \Phi} = \frac{1}{\|\nabla \Phi\|_g} df\cdot N,
\eqn
and, more generally, for any $j \geq 1$,
\bqn\label{partialPhij}
\frac{\partial^j f}{\partial \Phi^j} = \biggl(\frac{1}{\|\nabla \Phi\|_g} \iota_N d \biggr)^j f.
\eqn
A proof of this fact is given in Appendix B.

For $n = 3$, the conditions in (\ref{Fdeltaeqs}) are
\bqn \nb
&& F^{Im(3)}|_\Sigma = 0,
	\\ \nb
&&\biggl(3\frac{\partial F^{Im(3)}}{\partial \Phi} - F^{Im(2)}\biggr)\bigg|_\Sigma = 0,
	\\ \label{Fdeltaeqs2}
&&\biggl(3\frac{\partial^2 F^{Im(3)}}{\partial \Phi^2} - 2 \frac{\partial F^{Im(2)}}{\partial \Phi} 
+ F^{Im(1)}\biggr)\bigg|_\Sigma = 0,
	\\ \nb
&&\biggl(\frac{\partial^3 F^{Im(3)}}{\partial \Phi^3} - \frac{\partial^2 F^{Im(2)}}{\partial \Phi^2}  +\frac{\partial F^{Im(1)}}{\partial \Phi}
- F^{Im(0)}\biggr)\bigg|_\Sigma = 0. ~~
\eqn

\subsection{Distributional metric functions}

The field equations (\ref{eq1}) - (\ref{eq3}) involve second-order derivatives of the metric coefficients with respect to $t$ and sixth-order derivatives with respect to $x^i$. Thus, one might require that
the metric coefficients be $C^{1}$ with respect to $t$ and $C^{5}$ with respect to $x^{i}$, where $C^{n}$ indicates that the first $n$ derivatives exist and are continuous across the hypersurface
$\Phi = 0$.  However, this assumption eliminates the important case of an infinitely thin shell of matter supported on $\Sigma$. Therefore, we will instead make weaker assumptions, so that a thin shell
located on the hypersurface $\Phi = 0$ is in general allowed, and consider the case without a thin shell only as a particular case of our general treatment to be provided below. In fact, we shall impose the minimal requirement that {\em the corresponding problem be mathematically meaningful in terms of distribution theory}.
Then, in review of Eqs.(\ref{3.3b})-(\ref{3.3jb}), we find that the cases $\lambda = 1$ and $\lambda \not=1$ have different dependencies on the derivatives of $\mu$. In particular, the term $\mu'\mu''$ appears when $\lambda \not=1$. Thus, in the following we consider the two cases separately.

\subsubsection{$\lambda = 1$}

In this case, we assume that: (a) $\mu$ and $\nu$ are $C^5$ in each of the regions $M^+$ and $M^-$ up to the boundary $\Sigma$; (b) $\mu$ is $C^{0}$ across $\Sigma$; (c) $\nu$ is $C^{0}$ with respect to $t$ and $C^{2}$ with respect to $r$ across $\Sigma$.

The above regularity assumptions ensure that the mathematically ill-defined products $\delta(\Phi)^2$ and $\delta(\Phi)H(\Phi)$ do not appear in the field equations. Indeed, the terms in the field equations (\ref{3.3b}) - (\ref{3.3jb}) that could lead to products of this type are
\bq
\lb{5.5}
{\mu'}^{2},\;\; \dot{\mu}\mu',\;\; \dot{\nu}^2, \;\; {\nu''}^{2},\;\; \nu''\nu'''.
\eq
Our assumptions imply that these terms may contain $H(\Phi)^2$ but not $\delta(\Phi)^2$ or $\delta(\Phi)H(\Phi)$.

In order to compute the derivatives of $\mu$ and $\nu$, we note that
\bqn 
\mu & =& \mu^D = \mu^+ H(\Phi) + \mu^-[1-H(\Phi)], 
	 \nb\\
\nu &=& \nu^D = \nu^+ H(\Phi) + \nu^-[1-H(\Phi)],
\eqn
where the functions $\mu^+$ and $\nu^+$ are $C^5$ on $M^+$, while the functions $\mu^-$ and $\nu^-$ are $C^5$ on $M^-$. Let $V_\Sigma$ denote an open neighborhood of $\Sigma$. Let $\tilde{\mu}^+$ and $\tilde{\nu}^+$ denote $C^5$-extensions of $\mu^+$ and $\nu^+$ to $M^+ \cup V_\Sigma$. Let $\tilde{\mu}^-$ and $\tilde{\nu}^-$ denote $C^5$-extensions of $\mu^-$ and $\nu^-$ to $M^- \cup V_\Sigma$. Then the functions
\bqn\label{muhatnuhatdef}
\hat{\mu} \equiv \tilde{\mu}^+ - \tilde{\mu}^-,	 \qquad
\hat{\nu} \equiv \tilde{\nu}^+ - \tilde{\nu}^-,
\eqn
are defined on $V_\Sigma$ and the following relations are valid on $\Sigma$ whenever $\alpha + \beta \leq 5$:
\bqn
\lb{4.19}
\hat{\mu} = [\mu]^-, \quad 
\frac{\partial^{\alpha + \beta} }{\partial t^{\alpha} \partial r^{\beta}}\hat{\mu} = \biggl[\frac{\partial^{\alpha + \beta} }{\partial t^{\alpha} \partial r^{\beta}} \mu\biggr]^-, 
	\\ \label{muhatnuhat}
\hat{\nu} = [\nu]^-, \quad 
\frac{\partial^{\alpha + \beta} }{\partial t^{\alpha} \partial r^{\beta}}\hat{\nu} = \biggl[\frac{\partial^{\alpha + \beta} }{\partial t^{\alpha} \partial r^{\beta}} \nu\biggr]^-.
\eqn

Since $\mu$ is $C^0$ across $\Sigma$, we find
\bqn \nb
\mu_{,t} &=& (\mu_{,t})^D, 
	\\ \nb
 \mu_{,r} & = & (\mu_{,r})^D, 
	\\ \nb
\mu_{,tr} &=&  (\mu_{,tr})^{D} + \hat{\mu}_{,t} \delta(\Phi),
	\\ \nb
\mu_{,rt} &=&  (\mu_{,rt})^{D} - \dot{\cal{R}} \hat{\mu}_{,r}\delta(\Phi),
	\\ \nb
\mu_{,rr} &=&  (\mu_{,rr})^{D} + \hat{\mu}_{,r} \delta(\Phi).
	\\ \label{muderivatives}
\mu_{,rrr} &=&  (\mu_{,rrr})^{D} + 2\hat{\mu}_{,rr} \delta(\Phi) +  \hat{\mu}_{,r} \delta'(\Phi).
\eqn
Since $\mu$ is $C^0$ across $\Sigma$, the derivatives of $\mu^+$ and $\mu^-$ in any direction tangential to $\Sigma$ must 
coincide when evaluated on $\Sigma$. In particular, since the vector $U$ defined by
\bq
U^{\lambda} \equiv \delta^{\lambda}_{t} +  \dot{\cal{R}} \delta^{\lambda}_{r},
\eq
is tangential to $\Sigma$ (i.e. $U^{\lambda} N_{\lambda} = 0$), we obtain
$$
U^\lambda [\mu_{,\lambda}]^- = [\mu_{,t}]^- + \dot{\cal{R}} [\mu_{,r}]^- = 0,
$$
that is,
\bq
\hat{\mu}_{,t}  = -\dot{\cal{R}} \hat{\mu}_{,r},  
\eq
after Eq.(\ref{4.19}) is taken into account. Then,   from Eq.(\ref{muderivatives}) one finds  $\mu_{,tr} = \mu_{rt}$, as it is expected. 

Similarly, since $\nu$ is $C^0$ across $\Sigma$,
we also have
\bq
0 = U^\lambda [\nu_{,\lambda}]^- = [\nu_{,t}]^- + \dot{\cal{R}} [\nu_{,r}]^-.
\eq
But $[\nu_{,r}]^- = 0$, because $\nu$ is assumed to be $C^2$ with respect to $r$. Thus $[\nu_{,t}]^- = 0$. Therefore,  $\nu$ is in 
fact $C^1$ across $\Sigma$. The same argument applied to $\nu_{,t}$  and $\nu_{,r}$ now implies that $\nu$ is in fact $C^2$ across $\Sigma$.
We find 
\bqn \nb
\nu_{,t} &=& (\nu_{,t})^D, 
	\\ \nb
\nu_{,r} &=& (\nu_{,r})^{D}, 
	\\ \nb
\nu_{,rr} &=& (\nu_{,rr})^{D},
	\\ \nb
\nu^{(3)} &=& (\nu^{(3)})^D, 
	\\ \nb
\nu^{(4)} &=& (\nu^{(4)})^D + \hat{\nu}^{(3)} \delta(\Phi),
	\\ \label{nuderivatives}
\nu^{(5)} &=& (\nu^{(5)})^D + 2\hat{\nu}^{(4)} \delta(\Phi)
	 + \hat{\nu}^{(3)} \delta'(\Phi),
\eqn
where $\nu^{(n)} \equiv \partial^{n}\nu/\partial r^{n}$. 
We emphasize that the expressions on the right-hand sides of (\ref{muderivatives}) and (\ref{nuderivatives}) are independent of the extensions used to define $\hat{\mu}$ and $\hat{\nu}$ in (\ref{muhatnuhatdef}), because the values of $\hat{\mu}$, $\hat{\nu}$, and their partial derivatives of order $\leq 5$ are uniquely prescribed on $\Sigma$ in view of (\ref{muhatnuhat}).


We will find the junction conditions across $\Sigma$ by substituting the expressions (\ref{muderivatives}) and (\ref{nuderivatives}) for the derivatives of $\mu$ and $\nu$ into the field equations (\ref{3.3b}) - (\ref{3.3jb}).

\begin{table*}[htdp]
\lb{table1}
\begin{center}
\begin{tabular}{c|c|c|c|c}
Variation w.r.t. & Name of equation & General version & Spherically symmetric version & Associated junction condition \\ 
\hline
lapse $N(t)$ & Hamiltonian constraint & (\ref{eq1}) & (\ref{3.3b}) & (\ref{distHamiltonianconstraint}) \\
shift $N^i$ & Momentum constraint & (\ref{eq2}) & (\ref{3.3c}) & (\ref{vDJDJD}) \\
$\varphi$ & - & (\ref{eq4a}) & (\ref{3.3e}) & (\ref{vDJDJD}) \\
gauge field $A$ & - & (\ref{eq4b}) & (\ref{3.3f}) & (\ref{vDJDJD}) \\
metric $g_{ij}$ & Dynamical equations & (\ref{eq3}) & (\ref{3.3g}) and (\ref{3.3h}) & (\ref{distdynamic1}) and (\ref{distdynamic2}) \\
- & Energy conservation law & (\ref{eq5a}) & (\ref{3.3ja}) & (\ref{distenergyconservation}) \\
- & Momentum conservation law & (\ref{eq5b}) & (\ref{3.3jb}) & (\ref{distmomentumconservation}) \\
\end{tabular}
\end{center}
\label{default}
\caption{A list of all field equations for $\lambda = 1$.}
\end{table*}%

Suppose that the energy density $\rho_H = -2 J^{t}$ has the form
\bq
\rho_H = (\rho_H)^D + \sum_{n=0}^\infty \rho_H^{Im(n)} \delta^{(n)}(\Phi),
\eq
where it is understood that only finitely many of the $\rho_H^{Im(n)}$'s are nonzero.
Since, by (\ref{3.3a}),
$$\mathcal{L}_K = (\mathcal{L}_K)^D, \qquad \mathcal{L}_V = (\mathcal{L}_V)^D,$$
the Hamiltonian constraint (\ref{3.3b}) reads
\bqn \nb
&& \int_{r < \mathcal{R}(t)} \left( \mathcal{L}_K^- + \mathcal{L}_V^-  + 4 \pi G \rho_H^- \right) e^{\nu} r^2 dr
	\\ \label{distHamiltonianconstraint}
&& + \int_{r > \mathcal{R}(t)} \left( {\cal{L}}_{K}^+ + {\cal{L}}_{V}^+  + 4 \pi G \rho_H^+ \right) e^{\nu} r^{2} dr
	\\ \nb 
&& + 4\pi G \sum_{n=0}^\infty (-1)^n \frac{\partial^n}{\partial r^n}\bigg|_{r = \mathcal{R}(t)}\bigl(\rho_H^{Im(n)}e^{\nu} r^2\bigr)
= 0.
\eqn

The left-hand sides of Eqs.(\ref{3.3c}), (\ref{3.3e}) and (\ref{3.3f}) have no supports on the hypersurface  $r = {\cal{R}} (t)$. 
Thus, these equations remain unchanged in the regions $M^+$ and $M^-$, while on the hypersurface $\Sigma$ they yield
\bq \label{vDJDJD}
v = v^D, \qquad J_{\varphi} = (J_\varphi)^D, \qquad J_{A} = (J_A)^D.
\eq
In fact, in order to avoid that the ill-defined product $H(\Phi) \delta(\Phi)$ arises from the term $J_A A'$ in (\ref{3.3jb}), we will assume that $J_A$ is $C^0$.

The gauge field $A$ has dimension $[A] = 4$, so the action cannot contain terms like $A^n$ with $n \ge 2$, that is, it must be linear in $A$.
We therefore assume that $A$ has the form
\bqn\nb 
 A(t, r) &=& A^D + \sum_{n =0}^{\infty} A^{Im(n)}\delta^{(n)}(\Phi).
 \eqn
It follows that
 \bqn \nb
 A_{,r} &=& (A_{,r})^D + \bigl[\hat{A} + A_{,r}^{Im(0)}\bigr] \delta(\Phi)
 	\\ \nb
&& + \sum_{n=1}^\infty \bigl[ A_{,r}^{Im(n)} + A^{Im(n-1)}\bigr] \delta^{(n)}(\Phi),
	\\ \nb
 A_{,rr} &=& (A_{,rr})^D + \bigl[2\hat{A}_{,r} + A_{,rr}^{Im(0)}\bigr] \delta(\Phi)
 	\\ \nb
&& + \bigl[\hat{A} + 2A_{,r}^{Im(0)}+  A_{,rr}^{Im(1)}\bigr] \delta'(\Phi)
	\\ \nb
&& + \sum_{n=2}^\infty \bigl[ A_{,rr}^{Im(n)} + 2A_{,r}^{Im(n-1)} + A^{Im(n-2)}\bigr] \delta^{(n)}(\Phi).
\eqn
Thus,
\bqn 
F_{rr}^{A} &=& \frac{2}{r}\biggl\{(A_{,r})^D + \nu_{,r}A^D\nb\\
&& 
+ \bigl[\hat{A} + A_{,r}^{Im(0)} + \nu_{,r} A^{Im(0)}\bigr]\delta(\Phi) 
	\nb\\
&& + \sum_{n=1}^\infty \Bigl[\bigl(A_{,r}^{Im(n)} + A^{Im(n-1)}\bigr) \delta^{(n)}(\Phi)
	\nb\\
&& + \nu_{,r}  A^{Im(n)}\delta^{(n)}(\Phi)\Bigr] \biggr\},\nb\\
F_{\theta\theta}^{A} &=& (F_{\theta\theta}^A)^D + \sum_{n=0}^\infty F_{\theta\theta}^{A, Im (n)} \delta^{(n)}(\Phi), \nb
\eqn
where
\bqn \nb
(F_{\theta\theta}^A)^D &=& e^{-2\nu}\bigl[r^2(A_{,rr})^D 
 - \nu_{,r}r^2(A_{,r})^D +r(A_{,r})^D
 	\\ \nb
&& + r\nu_{,r} A^D -(1 - e^{2\nu})A^D\bigr],
	\\ \nb  
 F_{\theta\theta}^{A, Im (0)} &=& e^{-2\nu}\bigl[r^2(2\hat{A}_{,r} + A_{,rr}^{Im(0)}) \\ \nb
 && - r^2\nu_{,r}(\hat{A} + A_{,r}^{Im(0)}) 
 	\\ \nb
&& + r(\hat{A} + A_{,r}^{Im(0)})  + r\nu_{,r} A^{Im(0)} 
	\\ \nb
&& - (1 - e^{2\nu})A^{Im(0)}\bigr],
	\\ \nb
  F_{\theta\theta}^{A, Im (1)} &=& e^{-2\nu}\bigl[r^2(\hat{A} + 2 A_{,r}^{Im(0)} + A_{,rr}^{Im(1)}) 
  	\\ \nb
&&  - r^2 \nu_{,r}(A^{Im(0)} + A_{,r}^{Im(1)}) 
	\\ \nb
&&  + r(A^{Im(0)} + A_{,r}^{Im(1)}) 
  	\\ \nb
&& + r \nu_{,r} A^{Im(1)}   - (1 - e^{2\nu})A^{Im(1)}\bigr],
  	\\ \nb
   F_{\theta\theta}^{A, Im (n)} &=& e^{-2\nu}\bigl[r^2\big(A^{Im(n-2)} + 2 A_{,r}^{Im(n-1)} \\\ \nb
&& + A_{,rr}^{Im(n)}\big)  - r^2 \nu_{,r}(A^{Im(n-1)} + A_{,r}^{Im(n)}) 
	\\ \nb
&& + r(A^{Im(n-1)} + A_{,r}^{Im(n)}) 
	\\ \nb
&& + r \nu_{,r} A^{Im(n)}  - (1 - e^{2\nu})A^{Im(n)}\bigr], \; n \geq 2.
\eqn

From Eq.(\ref{A.4}) we find that the functions $\{F_n\}_{n=1}^6$ contain no delta functions whereas
\bqn \nb
  (F_7)_{rr} &=& (F_7)_{rr}^D - \frac{16e^{-4\nu}}{r^2}\hat{\nu}^{(3)}\delta(\Phi),
  	\\ \nb
  (F_8)_{rr} &=& (F_8)_{rr}^D - \frac{6e^{-4\nu}}{r^2} \hat{\nu}^{(3)}\delta(\Phi),
	\\ \nb
  (F_7)_{\theta\theta} &=& (F_7)_{\theta\theta}^D - 8re^{-6\nu}\bigl[(2\hat{\nu}^{(4)} - 16\nu_{,r}\hat{\nu}^{(3)}) \delta(\Phi)
  	\\ \nb
&&   + \hat{\nu}^{(3)} \delta'(\Phi)\bigr],
  	\\ \nb
  (F_8)_{\theta\theta} &=& (F_8)_{\theta\theta}^D - 3 r e^{-6\nu}\bigl[(2\hat{\nu}^{(4)} - 16 \nu_{,r} \hat{\nu}^{(3)})\delta(\Phi) 
  	\\ \nb
&&  + \hat{\nu}^{(3)}\delta'(\Phi)\bigr].
\eqn
Thus, (\ref{eq3a}) gives
\bqn \nb
F_{rr} &=& (F_{rr})^D - (16 g_7 + 6 g_8) \frac{e^{-4\nu}}{r^2 \zeta^4} \hat{\nu}^{(3)}\delta(\Phi),
	\\ \nb
F_{\theta\theta} &=& (F_{\theta\theta})^D - (8 g_7 + 3 g_8) \frac{r e^{-6\nu}}{\zeta^4}
	\\ \nb
&& \times \bigl[(2\hat{\nu}^{(4)} - 16 \nu_{,r} \hat{\nu}^{(3)})\delta(\Phi) + \hat{\nu}^{(3)}\delta'(\Phi)\bigr].
\eqn

Writing $p_r$ in the form 
\bqn\nb
p_r(t, r) = p_r^D + \sum_{n =0}^{\infty} p_r^{Im(n)}\delta^{(n)}(\Phi),
\eqn
we find that Eq.(\ref{3.3g}) remains unchanged in the regions $M^+$ and $M^-$, while on the hypersurface $\Sigma$ it yields
\bqn 
\label{distdynamic1}
\sum_{n=0}^\infty && \biggl\{\frac{e^{2\nu}}{r} [e^{-2\nu}(1 - 2r\nu') + \Lambda_g r^2 - 1]   A^{Im(n)} 
	\nb\\
&& + r\Bigl[F_{rr}^{Im(n)} + F_{rr}^{A Im(n)} + 8\pi G e^{2\nu} p_r^{Im(n)}\Bigr]\biggr\} \nb\\
&& ~~~~~~~~~ \times \delta^{(n)}(\Phi) = 0.
\eqn
Using (\ref{Fdeltaeqs}), Eq.(\ref{distdynamic1}) can be rewritten as a hierarchy of scalar equations on $\Sigma$.

Similarly, Eq.(\ref{3.3h}) remains unchanged in the regions $M^+$ and $M^-$, while on the hypersurface $\Sigma$ it yields 
\bqn 
 \label{distdynamic2}
&&  r (\hat{\mu}_{,r} e^{2\mu} + \dot{\mathcal{R}} \hat{\mu}_{,r} e^{\nu + \mu})\delta(\Phi)
	\nb\\
&& + \sum_{n=0}^\infty \biggl\{\frac{e^{2\nu}}{r} [e^{-2\nu}(1 - 2r\nu') + \Lambda_g r^2 - 1]   A^{Im(n)} 
   	 \nb\\
 &&  + \frac{e^{2\nu}}{r}\Bigl[ F_{\theta\theta}^{Im(n)} + F_{\theta\theta}^{AIm(n)} + 8\pi G r^2 p_\theta^{Im(n)}\Bigr]\biggr\} \delta^{(n)}(\Phi) = 0.\nb\\
 \eqn

Note that
\bqn \nb
\rho_{H,t} &=& (\rho_{H,t})^D + \bigl[\rho_{H,t}^{Im(0)} - \dot{\mathcal{R}} \hat{\rho}_{H} \bigr]\delta(\Phi) 
	\\ \nb
&& + \sum_{n=1}^\infty \bigl(\rho_{H,t}^{Im(n)} - \dot{\mathcal{R}} \rho_H^{Im(n-1)} \bigr) \delta^{(n)}(\Phi),
\eqn
and, by (\ref{vDJDJD}),
\bqn \nb
v_{,t} &=& (v_{,t})^D - \dot{\mathcal{R}} \hat{v} \delta(\Phi).
\eqn
Thus, in view of (\ref{vDJDJD}), the energy conservation law (\ref{3.3ja}) takes the form
\bqn \nb
 \int dr \, e^\nu r^2 \Bigl\{&&(\rho_{H,t})^D +\bigl[\rho_{H,t}^{Im(0)} - \dot{\mathcal{R}} \hat{\rho}_{H} \bigr]\delta(\Phi) 	
 	\\ \nb
&& + \sum_{n=1}^\infty \bigl(\rho_{H,t}^{Im(n)}  - \dot{\mathcal{R}} \rho_H^{Im(n-1)} \bigr) \delta^{(n)}(\Phi)
	\\ \nb
&& + \biggl(  (\rho_H)^D +  \sum_{n=0}^\infty \rho_H^{Im(n)} \delta^{(n)}(\Phi)
 	\\ \nb
&& + 4  (p_r)^D + 4 \sum_{n=0}^\infty p_r^{Im(n)} \delta^{(n)}(\Phi)\biggr)\nu_{,t}
	\\ \nb
 && + 4 ( (v_{,t})^D - \dot{\mathcal{R}} \hat{v} \delta(\Phi) - v^D \mu_{,t} ) 
 	\\ \nb
&& - 2\bigl( (J_{A,t})^D + \nu_{,t} (J_A)^D\bigr) \Bigr\} = 0,
\eqn
that is,
\bqn \nb
&& \biggl(\int_{r < \mathcal{R}(t)} + \int_{r > \mathcal{R}(t)}\biggr) e^\mu r^2 \bigl(\rho_{H,t} 
+ \nu_{,t}(\rho_H + 4 p_r)
 	\\ \nb
&&  + 4 v_{,t} - 4 v \mu_{,t}  
- 2(J_{A,t}+ \nu_{,t} J_A) \bigr) dr
 + \Bigl[e^\nu r^2\Bigl(\rho_{H,t}^{Im(0)} 
 	\\\nb
&& - \dot{\mathcal{R}} \hat{\rho}_{H} + \nu_{,t}\bigl(\rho_H^{Im(0)} + 4p_r^{Im(0)}\bigr)
 - 4\dot{\mathcal{R}} \hat{v}\Bigr)\Bigr]\Big|_{r = \mathcal{R}(t)}
	\\ \nb
&& + \sum_{n=1}^\infty (-1)^n \frac{\partial^n}{\partial r^n}\bigg|_{r = \mathcal{R}(t)}\Bigl[e^\nu r^2\Bigl(\rho_{H,t}^{Im(n)} - \dot{\mathcal{R}} \rho_H^{Im(n-1)} 
	\\  \label{distenergyconservation}
&& + \nu_{,t}\bigl(\rho_H^{Im(n)} + 4p_r^{Im(n)}\bigr)\Bigr)\Bigr] = 0.
 \eqn

The momentum conservation law (\ref{3.3jb}) remains unchanged in $M^+$ and $M^-$ while on the hypersurface $\Sigma$ it yields
\bqn \nb
&& - \hat{v} \delta(\Phi) + \hat{p}_r \delta(\Phi) 
	\\ \nb
&& + \sum_{n=0}^\infty \bigl[(p_r^{Im(n)})_{,r} \delta^{(n)}(\Phi) + p_r^{Im(n)} \delta^{(n+1)}(\Phi)\bigr] 
	\\ \nb
&& + \frac{2}{r}\sum_{n=0}^\infty ( p_r^{Im(n)} - p_\theta^{Im(n)}) \delta^{(n)}(\Phi)	
	\\ \nb
&& - \frac{1}{2}J_{A} \biggl[\bigl(\hat{A} + A_{,r}^{Im(0)}\bigr) \delta(\Phi)
 	\\ \nb
&& \qquad\qquad + \sum_{n=1}^\infty \bigl(A_{,r}^{Im(n)} + A^{Im(n-1)}\bigr) \delta^{(n)}(\Phi)\biggr]
	\\ \label{distmomentumconservation}
 && + e^{\nu-\mu} \dot{\mathcal{R}} \hat{v} \delta(\Phi)  = 0,
\eqn
where we have used that
\bqn \nb
p_{r}' = \hat{p}_r \delta(\Phi)  + \sum_{n=0}^\infty \bigl[(p_r^{Im(n)})_{,r} \delta^{(n)}(\Phi) + p_r^{Im(n)} \delta^{(n+1)}(\Phi)\bigr]. 
\eqn
This completes the general description of the junction conditions for the case $\lambda =1$, which are summarized in Table 1.

\subsubsection{$\lambda \not= 1$}

In this case,  the nonlinear terms
\bq
\lb{5.5b}
{\mu'}^{2},\;\; \dot{\mu}\mu',\;\; \mu'\mu'',\;\; \dot{\nu}^2, \;\; {\nu''}^{2},\;\; \nu''\nu''',
\eq
appear in the field equations (\ref{3.3b}) - (\ref{3.3jb}). Thus, to ensure these field equations are well-defined, we assume that:
(a) $\mu$ and $\nu$ are $C^5$ in each of the regions $M^+$ and $M^-$ up to the boundary $\Sigma$; (b) $\mu$ is 
$C^{0}$ with respect to $t$ and $C^{1}$ with respect to $r$ across $\Sigma$;  
(c) $\nu$ is $C^{0}$ with respect to $t$ and $C^{2}$ with respect to $r$ across $\Sigma$.

The same argument as above shows that $\nu$ is $C^2$ and that $\mu$ is $C^1$ across $\Sigma$. Equations (\ref{muderivatives}) and (\ref{nuderivatives}) for the derivatives of $\mu$ and $\nu$ are still valid, but since $\mu$ now is $C^1$, we have $\hat{\mu}_{,t} = \hat{\mu}_{,r} = 0$.
It follows that all the junction conditions (\ref{distHamiltonianconstraint}) - (\ref{distmomentumconservation}) remain unchanged, except that the presence of the term $\mu'''$ in (\ref{3.3e}) implies that the expression for $J_\varphi$ now may include a delta function:  
\bqn \label{Jvarphidelta}
&& J_{\varphi} = (J_\varphi)^D + (1-\lambda)\frac{e^{\mu - 3\nu}}{4 \pi G} \hat{\mu}_{,rr} \delta(\Phi).
\eqn

In what follows, we will consider some 
specific models of gravitational collapse for which the spacetime inside the collapsing sphere is described by the  Friedman-Lemaitre-Robertson-Walker (FLRW)
universe. 

{\section{Gravitational collapse of homogeneous and isotropic perfect fluid}}
\nequation

In this section, we consider the gravitational collapse of a spherical cloud consisting of a homogeneous and isotropic  perfect fluid \footnote{Gravitational collapse of a homogeneous and isotropic
dust fluid filled in the whole space-time was  considered in \cite{TS}, using a method proposed in \cite{WW09}.}, described 
by  the  FLRW universe,   
\bqn
ds^2 = -d\bar{t}^2 + a^2(\bar{t})\bigg(\frac{d\bar{r}^2}{1 - k\bar{r}^2} + \bar{r}^2 d^2\Omega\biggr),\nb
\eqn
where $k =0, \pm1$. Letting  $r = a(\bar{t})\bar{r}, \;  t = \bar{t}$, 
the corresponding ADM variables  take the form (\ref{3.1b}) with $N^- = 1$, and 
\bqn
\nu^{-}(t,r) &=& -\frac{1}{2} \ln\biggl(1 - k \frac{r^2}{a^2(t)}\biggr),\nb\\
\mu^{-}(t,r) &=&  \ln\biggl(\frac{-\dot{a}(t) r}{\sqrt{a^2(t) - k r^2}}\biggr),
\eqn
where $\dot{a} \le 0$ for a collapsing cloud. For a perfect fluid, we assume that 
\bq
\lb{PF}
p_\theta^{-} = p_r^{-} = p^-(t), \;\;\;  v = 0.  
\eq
We anticipate that the junction condition for $\nu$ requires  $k = 0$. Then, we find that
\bqn\label{numuflrwk0}
  \nu^{-}(t,r) = 0, \qquad \mu^{-}(t,r) = \ln\big({- rH}\big), \quad (k = 0), \quad
\eqn  
 where $ H \equiv \dot{a}(t)/a(t)$, and that
\bqn\nb
&&
\mathcal{L}_K^{-} = {3 (1-3 \lambda ) H^2}, \;\;\;  \mathcal{L}_V^{-} = 2\Lambda,
	\\  \label{FLRWk0Ls}
&& \mathcal{L}_\varphi^{-} =   \mathcal{L}_\lambda^{-} = 0,\;\;\; \mathcal{L}_A^{-} = 2 \Lambda_g A^{-}.
\eqn
It is easy to verify that the momentum constraint (\ref{3.3c}) is satisfied, whereas the equations (\ref{3.3e}) and (\ref{3.3f}) 
obtained by variation with respect to $\varphi$ and $A$ respectively, reduce to
\bqn
\lb{JJ}
  {3 \Lambda_g H}  +8 \pi  G J_\varphi^{-} &=& 0,
  	\\ \nb
  4\pi G J_A^{-} + \Lambda_g &=& 0.
\eqn 
Since $\nu^{-} = 0$, we have $F_{ij}^{-} = -\Lambda g_{ij}^{-}$, and the first dynamical equation (\ref{3.3g}) reduces to the condition
\bqn \nb
&& \frac{4}{r} a^2 A^{-}_{, r} + 2 a^2 \Lambda_g {A^{-}} +
2 (3 \lambda -1) a \ddot{a} + (3 \lambda -1) \dot{a}^2
	\\ \nb
&& ~~~~~~~~ + 2 a^2 (8 \pi  G p^- - \Lambda )  = 0.
 \eqn
If this condition is satisfied the second dynamical equation (\ref{3.3g}) also holds provided that
$A^{-}_{,r} - r A^{-}_{,rr} = 0.$
On the other hand, the momentum conservation law (\ref{3.3jb}) reduces to
$J_A^{-} A^{-}_{,r} = 0.$
We conclude that the general solution when $k = 0$ is given by
\bqn\label{flrwsolution1}
&&   J_\varphi^{-} = -\frac{3 \Lambda_g H}{8 \pi  G}, \quad
J_A^{-} = -\frac{\Lambda_g}{4\pi G},
\eqn
with $A^{-}= A^{-}(t)$ being given by 
\bqn
\lb{AA}
&&  \Lambda_g A^{-} +
 (3 \lambda -1)\left(\frac{\ddot{a}}{a} + \frac{H^2}{2}\right)
	   -     \Lambda = - 8 \pi  G p^-. ~~~~
\eqn

In the rest of this section, we consider only the case where $\Lambda_g =0$. Then, Eq.(\ref{flrwsolution1}) yields
\bqn\label{flrwsolution2}
&& J_A^{-} = J_\varphi^{-} = 0, 
\eqn
for which Eq.(\ref{AA}) shows that now $A^{-}(t)$    is an arbitrary function  of $t$, and $a(t)$ is given by
\bqn  
 \label{aeq}
&&  (3 \lambda -1)\left(\frac{\ddot{a}}{a} + \frac{H^2}{2}\right)
	   -     \Lambda = - 8 \pi  G p^-. 
\eqn

It is interesting to note that, since the Hamiltonian constraint is global, there is no analog of the Friedman equation in the current situation. This is in contrast 
to the case of HL cosmology \cite{HW}, where a Friedman-like equation still exists, because of the homogeneity and isotropy of the whole universe.\footnote{Considering that homogeneity and isotropy are good approximations for our observational  universe, this global Hamiltonian constraint
allows dust-like fluid to exist; this was first realized in \cite{Muka} where it was considered as a candidate of dark matter.} Although there is no analog of the Birkhoff theorem in HL theory, so that the spacetime outside the collapsing cloud can be either static or dynamical, 
we assume in this paper that the exterior solution is a static spherically symmetric vacuum spacetime. 
We also assume that the value of $\Lambda_g$ is the same in the exterior and interior regions, i.e. 
\bq
\lb{lambdag}
\Lambda_g^+ = \Lambda_g^- = 0.
\eq

It is convenient to consider the cases $\lambda = 1$ and $\lambda \not=1$ separately. 

\subsection{Gravitational Collapse with $\lambda = 1$}

We first consider the case of $\lambda = 1$. In this case, the static spherically symmetric exterior vacuum solution has the form  \cite{GSW}
\bqn
\lb{Outside}
&& \mu^+ = \mu^+(r)   = \frac{1}{2}\ln\biggl(\frac{2m^+}{r} + \frac{1}{3}\Lambda r^2 - 2A^+(r) \nb\\
&& ~~~~~~~~~~~~ ~~~~ ~~~~~~~ ~~~~~ + \frac{2}{r} \int_{r_0}^r A^+(r')dr'\biggr),\nb\\
&& \nu^+ = 0, 
\eqn
for which we find that
\bqn
 \label{exteriorsolution}
&& \mathcal{L}_K^+ = \frac{4}{r} A^+_{,r} - 2\Lambda, \quad \mathcal{L}_V^+ = 2\Lambda,   \quad \mathcal{L}_A^+ = 0,\nb\\
&& v^+ =   J_A^+ =   J_\varphi^+ =   \rho_H^+ = 0,
\eqn
where $m^+$, $r_0$ are constants and $A^+ = A^+(r)$ is a function  of  $r$ only, yet to be determined.  

As mentioned previously, the condition that $\nu$ be continuous across $\Sigma$ implies that $k = 0$.
We let the interior solution be of the form (\ref{numuflrwk0}), 
and assume that the thin shell of matter separating the interior and exterior solutions is such that
\bqn
\nb
&& p = p_{r}^{-}, \quad v = 0, \quad J_\varphi = J_\varphi^{Im(0)} \delta(\Phi),
	\\\nb
&& p_\theta = p_{\theta}^{-} + p_\theta^{Im(0)} \delta(\Phi), \quad \rho_H = \rho_H^- + \rho_H^{Im(0)}\delta(\Phi),\quad
	\\\label{shellassumptions}
&& A = A^D + A^{Im(0)}\delta(\Phi), \quad J_A = 0,  
\eqn
where 
$\rho^{+}_{H} =  J_\varphi^{\pm} = p_\theta^+ = p_r^+ = 0$.

\begin{proposition}\label{lambda1prop}
For the spacetime defined by (\ref{numuflrwk0}), (\ref{aeq}), (\ref{Outside}), the six 
junction conditions (\ref{distHamiltonianconstraint})-(\ref{distmomentumconservation}) reduce to the following six conditions:
\bqn \nb
&& \left(-6 H^2 + 2\Lambda + 4 \pi G \rho_H^-(t) \right) \frac{\mathcal{R}(t)^3}{3}
	\\ \label{JuncHamiltonianconstraint} 
&& \qquad  + 4\int_{\mathcal{R}(t)}^\infty  A_{,r}^+ r dr
 + 4\pi G \rho_H^{Im(0)} r^2 \Big|_{r = \mathcal{R}(t)} = 0,\nb\\
	\\[5pt] \label{JuncvDJDJD}
&& J_\varphi^{Im(0)} = 0,	
	\\[5pt] \label{Juncdynamic1}
&& \text{$A(t,r)$ is continuous across $\Sigma$},	
	\\[5pt] \label{Juncdynamic2}
&& A_{,t}^- = \mathcal{R} \biggl( \frac{\Lambda}{2} - H^2 \biggr)(\dot{\mathcal{R}} - H \mathcal{R}) - 8\pi G p_\theta^{Im(0)} H \mathcal{R},\nb\\
	\\[5pt]�\nb
 \label{Juncenergyconservation}
&& \rho_{H}^{Im(0)}(t, \mathcal{R}(t)) = e^{-\int_0^t \frac{2\dot{\mathcal{R}}(\tau)}{\mathcal{R}(\tau)}d\tau}\biggl[\rho_{H}^{Im(0)}(0, \mathcal{R}(0)) 
 	\\ \nb
&& + \int_0^t e^{\int_0^s \frac{2\dot{\mathcal{R}}(\tau)}{\mathcal{R}(\tau)}d\tau} \biggl(\frac{1}{4} H(s) \mathcal{R}(s)^2 \rho_{H,t}^-(s) \nb\\
&& ~~~~~~~~~~~~~~~~~~~~~~~ - \dot{\mathcal{R}}(s) \rho_H^-(s)\biggr) ds\biggr],
	\\[5pt]�\label{Juncmomentumconservation}
&& \text{$r p + 2p_\theta^{Im(0)}  = 0$ on $\Sigma$}.
\eqn
Moreover, the condition that $\mu$ be continuous across $\Sigma$ implies that
\bqn \label{mucontinuouscondition}
&&A^-_{,t} = \frac{\Lambda  - 3H^2}{2} \mathcal{R} \dot{\mathcal{R}}- HH_{,t} \mathcal{R}^2.
\eqn
\end{proposition}
\proofbegin
For the spacetime defined by (\ref{exteriorsolution})-(\ref{shellassumptions}), condition (\ref{distHamiltonianconstraint}) reduces to
\bqn\nb
&& \left(-6H(t)^2 + 2\Lambda + 4 \pi G \rho_H^-(t) \right) \int_0^{\mathcal{R}(t)}  r^2 dr
	\\ \nb 
&&  + 4\int_{\mathcal{R}(t)}^\infty  A_{,r}^+ r dr
 + 4\pi G \rho_H^{Im(0)} r^2 \Big|_{r = \mathcal{R}(t)}
= 0,
\eqn
which yields (\ref{JuncHamiltonianconstraint}). 
Moreover, condition (\ref{vDJDJD}) reduces immediately to (\ref{JuncvDJDJD}).

Conditions (\ref{distdynamic1}) and (\ref{distdynamic2}) reduce to
\bqn \label{distdynamic12}
&&  F_{rr}^{A Im(0)} \delta(\Phi) + F_{rr}^{A Im(1)} \delta'(\Phi) = 0,
\eqn
and
\bqn \nb
&& r (\hat{\mu}_{,r} e^{2\mu} + \dot{\mathcal{R}} \hat{\mu}_{,r} e^{\mu})\delta(\Phi)
	\\ \nb
&&+ \frac{1}{r} F_{\theta\theta}^{AIm(0)} \delta(\Phi) 
+ \frac{1}{r} F_{\theta\theta}^{AIm(1)} \delta'(\Phi) 
	\\ \label{distdynamic22}
&& + \frac{1}{r} F_{\theta\theta}^{AIm(2)}  \delta''(\Phi) + 8\pi G r p_\theta^{Im(0)} \delta(\Phi) = 0,
\eqn
respectively, where we have used that 
\bqn\label{FrrFthetatheta}
  F_{rr} = (F_{rr})^D, \qquad F_{\theta\theta} = (F_{\theta\theta})^D.
\eqn  
Now
\bqn \nb
F_{rr}^{A} &=& \frac{2}{r}\biggl\{(A_{,r})^D + \bigl[\hat{A} + A_{,r}^{Im(0)} \bigr]\delta(\Phi) \nb\\
&& ~~~~
	 + A^{Im(0)} \delta'(\Phi) \biggr\},
	 	\\ \label{FArrFAthetatheta}
  F_{\theta\theta}^{A} &=& (F_{\theta\theta}^A)^D + \sum_{n=0}^2 F_{\theta\theta}^{A, Im (n)} \delta^{(n)}(\Phi),
\eqn
where
\bqn \nb
(F_{\theta\theta}^A)^D &=& r^2(A_{,rr})^D +r(A_{,r})^D,
	\\ \nb  
 F_{\theta\theta}^{A, Im (0)} &=& r^2(2\hat{A}_{,r} + A_{,rr}^{Im(0)}) 
  + r(\hat{A} + A_{,r}^{Im(0)}),
	\\ \nb
  F_{\theta\theta}^{A, Im (1)} &=& r^2(\hat{A} + 2 A_{,r}^{Im(0)}) + rA^{Im(0)},
  	\\ \nb
   F_{\theta\theta}^{A, Im (2)} &=& r^2 A^{Im(0)}.
\eqn
Thus, equation (\ref{distdynamic12}) can be written as
\bqn\label{hatAArIm0}
\bigl[\hat{A} + A_{,r}^{Im(0)} \bigr]\delta(\Phi) + A^{Im(0)} \delta'(\Phi) = 0.
\eqn
{Thus, by}  (\ref{Fdeltaeqs2}), $A^{Im(0)}|_\Sigma = 0$. Hence, $A^{Im(0)}\delta(\Phi) = 0$ which gives 
$$0=(A^{Im(0)}\delta(\Phi))_{,r} = A_{,r}^{Im(0)} \delta(\Phi) + A^{Im(0)} \delta'(\Phi).$$
Equation (\ref{hatAArIm0}) then gives $\hat{A}|_\Sigma = 0$ so that in fact $A$ is continuous across $\Sigma$, which proves (\ref{Juncdynamic1}).
Equation (\ref{distdynamic22}) can now be written as
\bqn \nb
&& \Bigl[\hat{\mu}_{,r}  (e^{2\mu} + \dot{\mathcal{R}} e^{\mu})
+ 2\hat{A}_{,r} + 8\pi G p_\theta^{Im(0)} \Bigr]\delta(\Phi)  \nb\\
&& ~~~~~~~~~~~
 + \hat{A} \delta'(\Phi) = 0. 
\eqn
In view of (\ref{Fdeltaeqs2}) this yields
\bqn \label{hatmure2mu2}
\hat{\mu}_{,r}  (e^{2\mu} + \dot{\mathcal{R}} e^{\mu})
+ 2\hat{A}_{,r} + 8\pi G p_\theta^{Im(0)} = \frac{\partial \hat{A}}{\partial \Phi} \; \; \text{�on $\Sigma$}. \qquad
\eqn

Now observe that if a function $f(t,r)$ is $C^0$ across $\Sigma$, then
\bqn\label{dfhatdPhi}
  \frac{\partial \hat{f}}{\partial \Phi} = \frac{\partial \hat{f}}{\partial r} \; \text{ on $\Sigma$}.
\eqn  
Indeed, the continuity of $f$� implies that the derivative of $\hat{f}$ in any direction tangential to $\Sigma$ 
must vanish when evaluated on $\Sigma$; thus $\hat{f}_{,t} + \dot{\cal{R}} \hat{f}_{,r} = 0$ on $\Sigma$.
A computation using (\ref{gradPhicomponents}), (\ref{5.3a}), and (\ref{dfdPhi}) now gives (\ref{dfhatdPhi}).

On the other hand, since
$$\mu_{,r}^+ =  \frac{1}{2}(\Lambda r - 2A_{,r}^+) e^{-2\mu^+}, \qquad \mu_{,r}^- = \frac{1}{r},$$
we find
\bqn\label{hatmur}
\hat{\mu}_{,r} = \frac{1}{2}(\Lambda r - 2A_{,r}^+) e^{-2\mu^+} - \frac{1}{r}.
\eqn

Inserting the equations (\ref{dfhatdPhi}) and (\ref{hatmur}) into (\ref{hatmure2mu2}), we find
\bqn \nb
&& \biggl(\frac{1}{2}(\Lambda r - 2 A_{,r}^+)e^{-2\mu} - \frac{1}{r}\biggr)(e^{2\mu} + \dot{\mathcal{R}}e^\mu)  + \hat{A}_{,r} 
	\\ \nb
&&\qquad + 8\pi G p_\theta^{Im(0)} = 0 \text{  on $\Sigma$.}
\eqn
Since $ \hat{A}_{,r} = A_{,r}^+ = A_{,t}^- \dot{\mathcal{R}}^{-1}$, simplification yields (\ref{Juncdynamic2}).

Condition (\ref{distenergyconservation}) reduces to
\bqn \nb
&&\int_0^{\mathcal{R}(t)} e^\mu r^2 \rho_{H,t}^- dr
+ r^2 \Bigl[ \rho_{H,t}^{Im(0)} - \dot{\mathcal{R}} \hat{\rho}_H \Bigr]\Big|_{r = \mathcal{R}(t)}
	\\
&& + \frac{\partial}{\partial r}\bigg|_{r = \mathcal{R}(t)} \Bigl[r^2 \dot{\mathcal{R}} \rho_H^{Im(0)}\Bigr] = 0.
 \eqn
That is,
 \bqn \nb
&&- \frac{\dot{a}(t)\rho_{H,t}^-(t)}{a(t)}   \int_0^{\mathcal{R}(t)} r^3  dr 
	\\ \nb
&& + \mathcal{R}(t)^2 \Bigl[\rho_{H,t}^{Im(0)}(t, \mathcal{R}(t)) + \dot{\mathcal{R}}(t) \rho_H^-(t)\Bigr]
	\\ \nb
&& 
 + 2\mathcal{R}(t) \dot{\mathcal{R}}(t) \rho_H^{Im(0)} (t, \mathcal{R}(t))
 	\\
&& + \mathcal{R}(t)^2 \dot{\mathcal{R}}(t)\rho_{H,r}^{Im(0)}(t, \mathcal{R}(t)) = 0.
 \eqn
Consequently, 
\bqn \nb
&&-\frac{H(t) \rho_{H,t}^-(t) \mathcal{R}(t)^2}{4}  + \frac{d}{dt}\Bigl[\rho_{H}^{Im(0)}(t, \mathcal{R}(t))\Bigr]
	\\ \label{preJuncenergyconservation}
&&  +  2 \frac{\dot{\mathcal{R}}(t)}{\mathcal{R}(t)} \rho_H^{Im(0)} (t, \mathcal{R}(t)) + \dot{\mathcal{R}}(t) \rho_H^-(t) = 0.
\eqn
Solving this differential equation for $\rho_H^{Im(0)}$, we find (\ref{Juncenergyconservation}).

Condition (\ref{distmomentumconservation}) reduces to 
 \bqn \nb
&&  \left(p   + \frac{2}{r} p_\theta^{Im(0)}\right) \delta(\Phi) = 0.
\eqn
This yields (\ref{Juncmomentumconservation}). 

Finally, the condition that $\mu$ be continuous across $\Sigma$ can be written as
\bqn \nb
&&\frac{2m^+}{\mathcal{R}(t)} + \frac{1}{3}\Lambda \mathcal{R}(t)^2 - 2A^+(\mathcal{R}(t)) 
	\\ \label{premucondition}
&&+ \frac{2}{\mathcal{R}(t)} \int_{r_0}^{\mathcal{R}(t)} A^+(r')dr'
= H^2\mathcal{R}^2.
\eqn
Since $A$ is continuous across $\Sigma$, we have $A^+(\mathcal{R}(t)) = A^-(t)$. Hence, multiplying 
(\ref{premucondition}) by $\mathcal{R}$ and then differentiating with respect to $t$, we find
\bqn \nb
&&\Lambda \mathcal{R}^2 \dot{\mathcal{R}} 
- 2A^-_{,t} \mathcal{R} = 2HH_{,t} \mathcal{R}^3 + 3H^2\mathcal{R}^2\dot{\mathcal{R}}.
\eqn
Solving this equation for $A^-_{,t}$, we find (\ref{mucontinuouscondition}).
\proofend

The conditions (\ref{Juncdynamic2}) and (\ref{mucontinuouscondition}) imply that
\bqn \nb
&& \biggl( \frac{\Lambda}{2} - H^2 \biggr)(\dot{\mathcal{R}} - H \mathcal{R}) - 8\pi G p_\theta^{Im(0)} H 
	\\ \nb
&& \hspace{3cm} = \frac{\Lambda  - 3H^2}{2} \dot{\mathcal{R}}- HH_{,t} \mathcal{R},
\eqn
i.e.
\bqn \nb
&& H \dot{\mathcal{R}} + (2H^2 + 2H_{,t} - \Lambda ) \mathcal{R}
 - 16\pi G p_\theta^{Im(0)}  = 0.
\eqn
Solving this equation for $\mathcal{R}(t)$ we find the following equation which expresses $\mathcal{R}(t)$ in terms of $H(t)$ and the pressure $p_\theta^{Im(0)}$ on the shell:
\bqn \nb
\mathcal{R}(t) & = & e^{-\int_0^t I(s) ds}\biggl\{\mathcal{R}(0) 
	\\ \label{Rsolvedfor}
&& + 16\pi G \int_0^t e^{\int_0^s I(\tau) d\tau}\frac{p_\theta^{Im(0)}(s, \mathcal{R}(s))}{H(s)} ds\biggr\}, \qquad \quad
\eqn
where $I(t)$ is defined by
\bqn\label{Idef}
 I = 2H + \frac{2H_{,t}}{H} - \frac{\Lambda}{H}.
\eqn

\subsection{Dust Collapse with $\lambda = 1$}

Suppose now that the perfect fluid in the interior region consists of dust, i.e. 
\bqn\label{przero}
p_r^-  = p^{-}_{\theta} = 0.
\eqn 
Then, the condition (\ref{Juncmomentumconservation}) implies that
\bqn
  p_\theta^{Im(0)} = 0.
\eqn  
Solving equation (\ref{aeq}) for $a(t)$ we find
\bqn \label{acases}
a(t) = \cases{a_0 \cosh^{\frac{2}{3}}\biggl(\frac{\sqrt{3\Lambda}}{2} (t - t_0)  \biggr), & $\Lambda \not= 0$,\cr
a_0   (t_0 - t)^{2/3}, & $\Lambda = 0$,\cr}
\eqn
where $a_0$ and $t_0$ are constants. 
In the following, let us consider the cases $\Lambda \not= 0$ and $\Lambda = 0$, separately.

\subsubsection{$\Lambda > 0$}

In this case, substituting the expression for $a(t)$ into (\ref{Idef}) we obtain
\bqn\nb
  I(t) =  \sqrt{\frac{\Lambda}{3}}\tanh\biggl(\frac{\sqrt{3\Lambda}}{2} (t_0  - t)  \biggr),
\eqn
and then (\ref{Rsolvedfor}) yields
\bqn\label{Rdustformula}
  \mathcal{R}(t) = \mathcal{R}_0 \cosh^{\frac{2}{3}}\biggl(\frac{\sqrt{3\Lambda}}{2} (t_0  - t)\biggr), 
\eqn
where $\mathcal{R}_0$ is a constant.
Condition (\ref{mucontinuouscondition}) now implies that $A^-_{,t} = 0$, i.e. $A^-(t) = A_0$ for some constant $A_0$. 
Then, by (\ref{Juncdynamic1}), $A^+(\mathcal{R}(t)) = A_0$. That is, $A^+(r) = A_0$ for all $r$ such that $r = \mathcal{R}(t)$ for some $t$. 
Hence, the form of (\ref{Rdustformula}) implies that $A^+ = A_0$ for all $(t,r)$ in the exterior region. This gives
\bqn  
  A(t,r) = A_0.
\eqn
Condition (\ref{JuncHamiltonianconstraint}) now implies
\bqn  \nb
 &&  \rho_H^{Im(0)}(t, \mathcal{R}(t)) = -\frac{(-6H^2 + 2\Lambda + 4\pi G \rho_H^-(t))\mathcal{R}(t)}{12\pi G}
  	\\  \label{rhoHIm0dust}
&& = - \mathcal{R}_0 \frac{\Lambda + \pi G [1 + \cosh(\sqrt{3\Lambda}(t_0  - t))]\rho_H^-(t)}{6\pi G \cosh^{\frac{4}{3}}(\frac{\sqrt{3\Lambda}}{2}(t_0  - t))}.
\eqn
Substituting this into condition (\ref{Juncenergyconservation}), or its equivalent form (\ref{preJuncenergyconservation}), we infer that $\rho_H^-(t)$ satisfies:
\bqn \nb
&& -\frac{\mathcal{R}_0}{12}\cosh^{\frac{1}{3}}\biggl(\frac{\sqrt{3\Lambda}}{2} (t_0  - t)  \biggr)
\biggl\{4 \cosh^{\frac{1}{3}}\biggl(\frac{\sqrt{3\Lambda}}{2} (t_0  - t)  \biggr)
	\\ \nb
&& - \mathcal{R}_0\sqrt{3\Lambda} \sinh\biggl(\frac{\sqrt{3\Lambda}}{2} (t_0  - t)  \biggr)\biggr\}
\rho_{H,t}^-(t)  = 0,
\eqn
i.e.
\bqn
\rho_H^-(t) = \rho_{H}^{(0)}, 
\eqn
where $\rho_{H}^{(0)}$ is a constant.
All the conditions of Proposition \ref{lambda1prop} are now satisfied. It only remains to consider the condition that $\mu$ be continuous across $\Sigma$. This condition reduces to
\bqn \nb
0 &=& \frac{2m^+}{\mathcal{R}} + \frac{1}{3}\Lambda \mathcal{R}^2 - 2A_0 + \frac{2}{\mathcal{R}} A_0(\mathcal{R} - r_0)
-  H^2 \mathcal{R}^2
	\\ \nb
&=& \frac{6 m^+ - 6A_0 r_0 + \mathcal{R}_0^3 \Lambda}{3\mathcal{R}_0 \cosh^{2/3}(\frac{\sqrt{3\Lambda}}{2} (t_0  - t))}.
\eqn
That is, the parameter $r_0$ is fixed by
\bqn 
  r_0 = \frac{6 m^+  + \mathcal{R}_0^3 \Lambda}{6A_0}.
\eqn
This implies that
\bqn
\mu^+ = \frac{1}{2}\ln\biggl(\frac{\Lambda r^2}{3} - \frac{\Lambda \mathcal{R}_0^3}{3r}\biggr).
\eqn
Since all the field equations and junction conditions are now satisfied we have proved the following result.

\begin{proposition}
Ho\v{r}ava-Lifshitz gravity admits the following explicit solution when $\lambda = 1$ and $\Lambda > 0$:
\bqn 
\lb{PropV.2}
&& \mu^+ = \frac{1}{2}\ln\biggl(\frac{\Lambda r^2}{3} - \frac{{\cal{R}}_{0}^3\Lambda}{3r}\biggr), \quad
\mu^- = \ln\big(-H(t)r\big),
	 \nb\\
&&  \nu = 0, \quad H(t) = - \sqrt{\frac{\Lambda}{3}}\tanh\biggl(\frac{\sqrt{3\Lambda}}{2} (t_0 - t)  \biggr),
	 \nb\\
&& \mathcal{R}(t) = \mathcal{R}_0 \cosh^{\frac{2}{3}}\biggl(\frac{\sqrt{3\Lambda}}{2} (t_0 - t)  \biggr),
	 \\
&& p_r = p_\theta = 0, \quad \rho_H^-(t) = \rho_{H}^{(0)}, \quad A(t,r) = A_0,
	 \nb\\
&& \text{$\rho_H^{Im(0)}$ is given by (\ref{rhoHIm0dust})},\nb
\eqn
where $t_0$, $\mathcal{R}_0$, $A_0$, and $\rho_{H}^{(0)}$ are constants.
\end{proposition}

\begin{figure}[tbp]
\centering
\includegraphics[width=8cm]{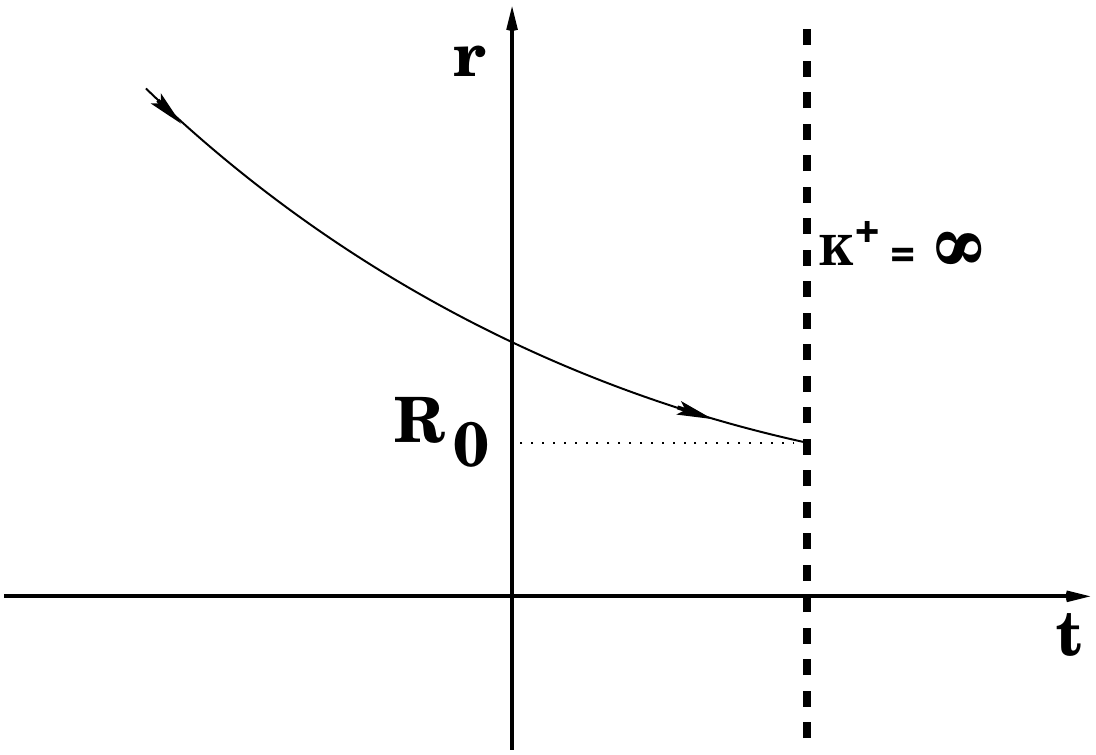} 
\caption{The evolution of the surface of the collapsing star for $\lambda = 1$ and $\Lambda > 0$, given by Eq.(\ref{PropV.2}). At the moment
$t = t_0$,  the star collapses to its minimal radius   ${\cal{R}}(t_0) =  {\cal{R}}_0$, at which the extrinsic curvature $K^{+}$ becomes unbounded,
while the four-dimensional Ricci scalar remains finite. }
\label{fig2}
\end{figure}

For $t < t_0$ the dust cloud is contracting.
As $t \nearrow t_0$, the radius of the dust sphere approaches its minimal value of $\mathcal{R} = \mathcal{R}_0$ at $t = t_{0}$, and the function
 $e^{\mu^+}$  approaches zero:
$$
\lim_{t \nearrow t_0} \mathcal{R}(t) = \mathcal{R}_0, \qquad
\lim_{t \nearrow t_0} e^{\mu^+(t)} = 0,
$$
as shown schematically in Fig. \ref{fig2}.   After the star collapses to this point, it is not clear how spacetime
{evolutes}, because $\mu_{+}$ 
becomes unbounded as one can see from Eq.(\ref{PropV.2}), for which the extrinsic scalar $K^+$,
\bq
\lb{K}
K^+(r) =    e^{\mu^+(r)}\left({\mu^+_{,r}(r)} + \frac{2}{r}\right),
\eq
also becomes unbounded, which indicates the existence of a scalar singularity at this point \cite{CW}.  However, such a singularity is weak. In particular,   the corresponding four-dimensional 
Ricci scalar remains finite, ${}^{(4)}R = 4 \Lambda$.  Thus, it is not clear whether the spacetime across this point is extendable or not. 

In addition,  Eq.(\ref{rhoHIm0dust}) shows that $\rho_H^{Im(0)}$ and $\rho_H^-$ cannot both be positive. To understand this,  
letting $M = - \Lambda \mathcal{R}_0^3/6$ we can write $\mu^{+}$ in the form,
\bq
\mu^+ = \frac{1}{2}\ln\biggl(\frac{2M}{r} + \frac{\Lambda r^2}{3}\biggr).
\eq
However,  this is nothing but the Schwarzschild-de Sitter solution with mass $M$ and a cosmological constant $\Lambda$, where
$M$ is  negative.

\subsubsection{$\Lambda < 0$}

In this case, substituting the expression for $a(t)$ into (\ref{Idef}) we obtain
\bqn\nb
  I(t) = \sqrt{\frac{|\Lambda|}{3}}\tan\biggl(\frac{\sqrt{3|\Lambda|}}{2} (t - t_0)  \biggr),
\eqn
and then (\ref{Rsolvedfor}) yields
\bqn
  \mathcal{R}(t) = \mathcal{R}_0 \cos^{\frac{2}{3}}\biggl(\frac{\sqrt{3|\Lambda|}}{2} (t - t_0)\biggr), 
\eqn
where $\mathcal{R}_0$ is another constant.
Condition (\ref{mucontinuouscondition}) now implies that $A^-_{,t} = 0$, i.e. $A^-(t) = A_0$ for some constant $A_0$. 
Then, by (\ref{Juncdynamic1}), $A^+(\mathcal{R}(t)) = A_0$. That is, $A^+(r) = A_0$ for all $r$ such that $r = \mathcal{R}(t)$ for some $t$. 
We will assume that $A^+ = A_0$ for all $(t,r)$ in the exterior region, i.e.
$
  A(t,r) = A_0.
$
Condition (\ref{JuncHamiltonianconstraint}) now implies
\bqn  \nb
 &&  \rho_H^{Im(0)}(t, \mathcal{R}(t)) = -\frac{(-6H^2 + 2\Lambda + 4\pi G \rho_H^-(t))\mathcal{R}(t)}{12\pi G}
  	\\  \label{rhoHIm0dust2}
&& = \mathcal{R}_0 \frac{|\Lambda| - \pi G [1 + \cos(\sqrt{3|\Lambda|}(t-t_0))]\rho_H^-(t)}{6\pi G \cos^{\frac{4}{3}}(\frac{\sqrt{3|\Lambda|}}{2}(t-t_0))}.
\eqn
Substituting this into condition (\ref{Juncenergyconservation}), or its equivalent form (\ref{preJuncenergyconservation}), we infer that $\rho_H^-(t)$ satisfies
\bqn
\rho_H^-(t) = \rho_{H}^{(0)}, 
\eqn
where $\rho_{H}^{(0)}$ is a constant.
All the conditions of Proposition \ref{lambda1prop} are now satisfied, while  the condition that $\mu$ be continuous across $\Sigma$   reduces to
\bqn \nb
0 &=& \frac{2m^+}{\mathcal{R}} + \frac{1}{3}\Lambda \mathcal{R}^2 - 2A_0 + \frac{2}{\mathcal{R}} A_0(\mathcal{R} - r_0)
-  H^2 \mathcal{R}^2
	\\ \nb
&=& \frac{6 m^+ - 6A_0 r_0 + \mathcal{R}_0^3 \Lambda}{3\mathcal{R}_0 \cos^{2/3}(\frac{\sqrt{3|\Lambda|}}{2} (t - t_0))}.
\eqn
Thus, the parameter $r_0$ is fixed by
\bqn 
  r_0 = \frac{6 m^+  + \mathcal{R}_0^3 \Lambda}{6A_0}.
\eqn
This implies that
\bqn
\mu^+ = \frac{1}{2}\ln\biggl(\frac{2M}{r} -\frac{|\Lambda|}{3}  r^2\biggr),
\eqn
where $M \equiv |\Lambda| \mathcal{R}_0^3/6$. Clearly, this corresponds to the Schwarzschild-anti-de Sitter solution. For $\mu^{+}$ to be real, we must assume that
$r \le {\cal{R}}_0$. Similar to the last case, the extrinsic curvature $K^+$ at $r =  {\cal{R}}_0$ becomes unbounded, while the four-dimensional Ricci scalar ${}^{(4)}R$ remains
constant. Thus, in this case it is also not clear whether or not the spacetime is extendable cross $r =  {\cal{R}}_0$. 

In any case,  all the field equations and junction conditions are now satisfied for $r \le {\cal{R}}_0$, and  we have proved the following result.

\begin{proposition}
Ho\v{r}ava-Lifshitz gravity admits the following explicit solution when $\lambda = 1$ and $\Lambda < 0$:
\bqn 
\lb{PropV.3}
&& \mu^+ = \frac{1}{2}\ln\biggl(\frac{|\Lambda|}{3 r}(\mathcal{R}_0^3 - r^3)\biggr), \quad
\mu^- = \ln(-H(t)r),
	 \nb\\
&&  \nu = 0, \quad H(t) = -\sqrt{\frac{|\Lambda|}{3}}\tan\biggl(\frac{\sqrt{3|\Lambda|}}{2} (t - t_0)  \biggr),
	\nb\\ 
&& \mathcal{R}(t) = \mathcal{R}_0 \cos^{\frac{2}{3}}\biggl(\frac{\sqrt{3|\Lambda|}}{2} (t - t_0)  \biggr),
	\\ 
&& p_r = p_\theta = 0, \quad \rho_H^-(t) = \rho_{H}^{(0)}, \quad A(t,r) = A_0,
	 \nb\\
&& \text{$\rho_H^{Im(0)}$ is given by (\ref{rhoHIm0dust2})},\nb
\eqn
where $t_0$, $\mathcal{R}_0$, $A_0$, and $\rho_{H}^{(0)}$ are constants.  
\end{proposition}

\begin{figure}[tbp]
\centering
\includegraphics[width=8cm]{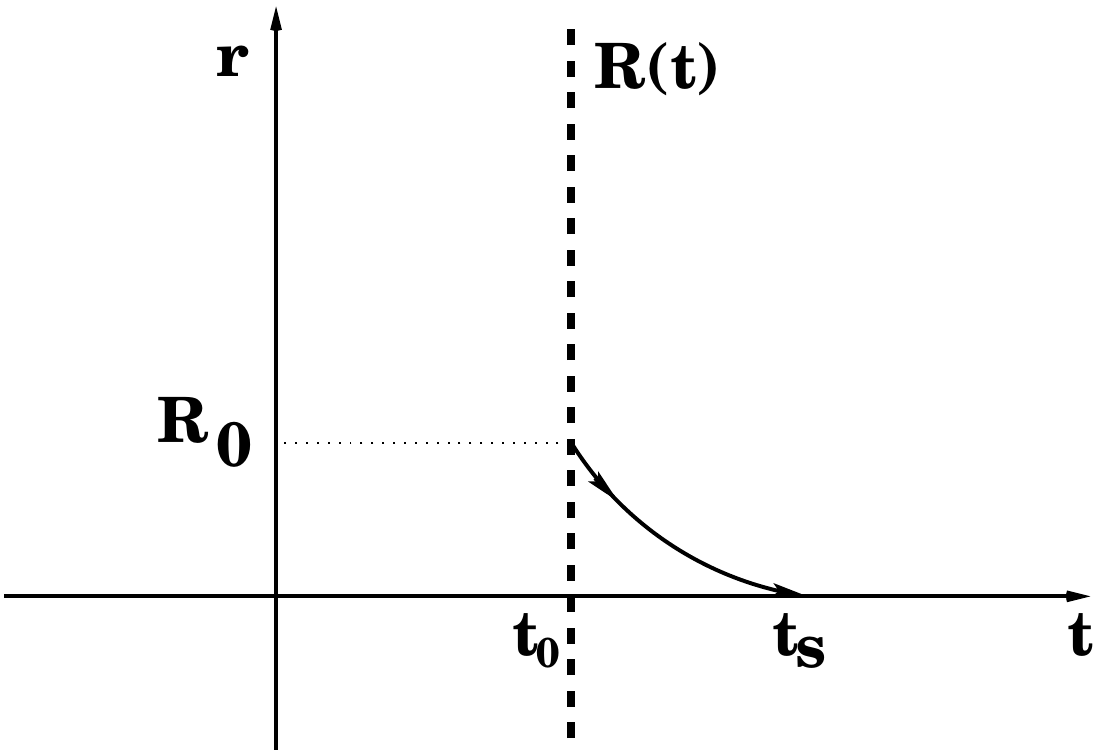} 
\caption{The evolution of the surface of the collapsing star for $\lambda = 1$ and $\Lambda < 0$, given by Eq.(\ref{PropV.3}). The star starts to collapse at a time $t = t_i \ge t_0$. At the later time $t = t_s$, at which ${\cal{R}}(t_s) =  0$, the star collapses and a central singularity is formed.}
\label{fig3}
\end{figure}

The evolution of the surface of the collapsing star is illustrated in Fig. \ref{fig3}. The collapse starts at an initial time $t_i \le t_0$, and at time $t = t_s$, the star collapses to a central singularity at which we have $\mathcal{R}(t_s) = 0$, where $t_s \equiv t_0 + \pi/\sqrt{3|\Lambda|}$.
Equation (\ref{rhoHIm0dust2}) shows that now both $\rho_H^{Im(0)}$ and $\rho_H^{-}$ can be positive, provided that $|\Lambda| > 2 \pi G \rho_H^{(0)}.$


\subsubsection{$\Lambda = 0$}

In this case, substituting the expression (\ref{acases}) for $a(t)$ into (\ref{Idef}) we obtain
\bqn\nb
  I(t) = \frac{2}{3(t_0 -t)},
\eqn
and then (\ref{Rsolvedfor}) yields
\bqn\label{Rdustformula2}
  \mathcal{R}(t) = \mathcal{R}_0 (t_0 -t)^{\frac{2}{3}}, 
\eqn
where $\mathcal{R}_0$ is a constant.
Condition (\ref{mucontinuouscondition}) now implies that $A^-_{,t} = 0$, i.e. $A^-(t) = A_0$ for some constant $A_0$. 
Then, by (\ref{Juncdynamic1}), $A^+(\mathcal{R}(t)) = A_0$. That is, $A^+(r) = A_0$ for all $r$ such that $r = \mathcal{R}(t)$ for some $t$. 
Hence (\ref{Rdustformula2}) implies that $A^+ = A_0$ for all $(t,r)$ in the exterior region. Thus, in the present case we also have  
$  A(t,r) = A_0.
$
Condition (\ref{JuncHamiltonianconstraint}) now implies
\bqn  \nb
 &&  \rho_H^{Im(0)}(t, \mathcal{R}(t)) = -\frac{(-6H^2 + 4\pi G \rho_H^-(t))\mathcal{R}(t)}{12\pi G}
  	\\  \nb
&& = \mathcal{R}_0 \frac{2 - 3G\pi(t_0-t)^2 \rho_H^-(t)}{9G\pi(t_0 -t)^{4/3}}.
\eqn
Substituting this into condition (\ref{Juncenergyconservation}), or its equivalent form (\ref{preJuncenergyconservation}), we infer that 
\bqn
\rho_H^-(t) = \rho_{H}^{(0)}, 
\eqn
where $\rho_{H}^{(0)}$ is a constant.
All the conditions of Proposition \ref{lambda1prop} are now satisfied, and  the condition that $\mu$ be continuous across $\Sigma$ becomes
\bqn \nb
0 &=& \frac{2m^+}{\mathcal{R}} - 2A_0 + \frac{2}{\mathcal{R}} A_0(\mathcal{R} - r_0)
-  H^2 \mathcal{R}^2
	\\ \nb
&=& 2\frac{9m^+ - 9A_0 r_0 - 2\mathcal{R}_0^3}{9\mathcal{R}(t)}.
\eqn
Hence,  the parameter $r_0$ is fixed to
\bqn 
  r_0 = \frac{9 m^+ - 2\mathcal{R}_0^3}{9A_0},
\eqn
which implies that
\bqn
  \mu^+ = \frac{1}{2}\ln\biggl(\frac{r_g}{r}\biggr),\;\; \nu^{+} = 0,\;\; N^{+} = 1, 
\eqn
where $r_g \equiv 4\mathcal{R}_0^3/9$. This is nothing but is the Schwarzschild solution written in the Painlev\'e-Gullstrand coordinates \cite{GP}. 
All the field equations and junction conditions are satisfied, so we have proved the following result.

\begin{proposition}
Ho\v{r}ava-Lifshitz gravity admits the following explicit solution when $\lambda = 1$ and $\Lambda = 0$:
\bqn 
\lb{PropV.4}
&& \mu^+ =  \frac{1}{2}\ln\biggl(\frac{r_g}{r}\biggr), \quad
\mu^- = \ln\big(-H(t)r\big),
	 \nb\\
&&  \nu = 0, \quad H(t) = -\frac{2}{3(t_0 -t)},
	\nb\\
&& \mathcal{R}(t) = \mathcal{R}_0 (t_0 -t)^{\frac{2}{3}},
	\\ 
&& p_r = p_\theta = 0, \quad \rho_H^-(t) = \rho_{H}^{(0)}, \quad A(t,r) = A_0,
	 \nb\\
&& \rho_H^{Im(0)} = \mathcal{R}_0 \frac{2 - 3G\pi(t_0-t)^2 \rho_{H}^{(0)}}{9G\pi(t_0 -t)^{4/3}},\nb
\eqn
where $t_0$, $\mathcal{R}_0$, $A_0$, and $\rho_{H}^{(0)}$ are constants.  
\end{proposition}

\begin{figure}[tbp]
\centering
\includegraphics[width=8cm]{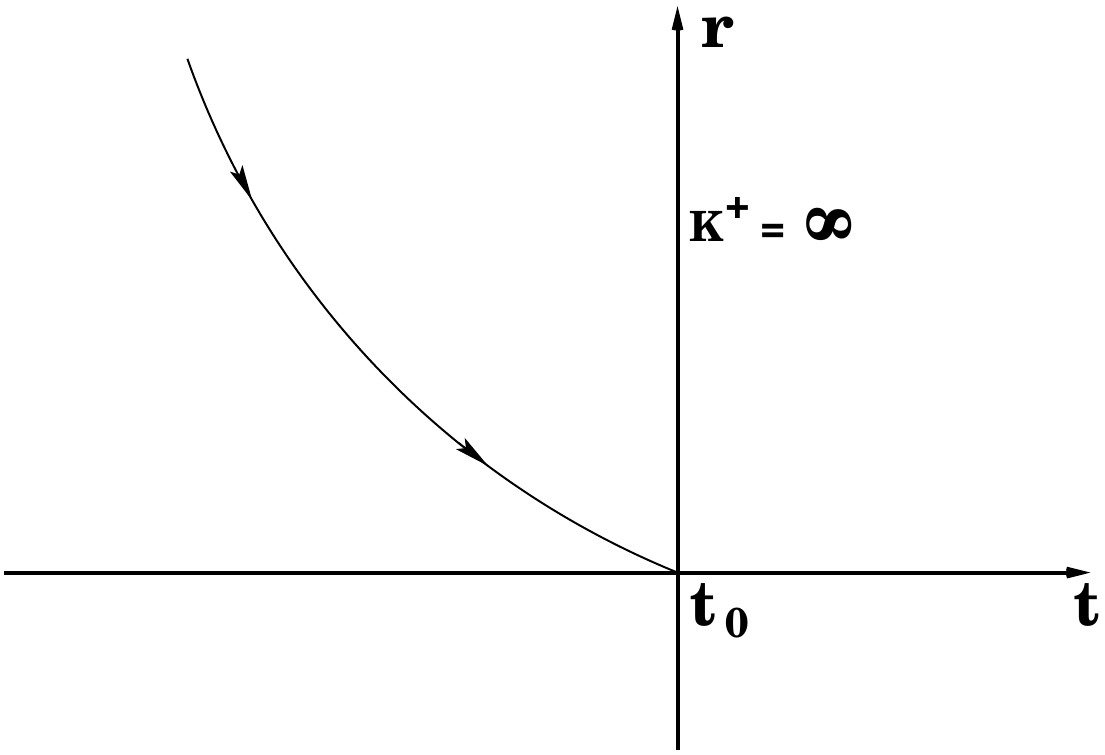} 
\caption{The evolution of the surface of the collapsing star for $\lambda = 1$ and $\Lambda = 0$, given by Eq.(\ref{PropV.4}). At the moment
$t = t_i \le t_0$,  the star starts to collapse until the moment $t = t_0$, at which we have    ${\cal{R}}(t_0) =  0$, whereby a central singularity is formed.}
\label{fig4}
\end{figure}

The evolution of the surface of the collapsing star is shown in Fig. \ref{fig4}. The star begins to collapse at the moment $t_i$ with a radius 
$\mathcal{R}_i [\equiv \mathcal{R}(t_i)]$ until the moment $t = t_0$, at which we have ${\cal{R}}(t_0) =  0$ and a central singularity is formed. 
The spacetime outside of the star is given by the Schwarzschild solution.   
Thus, as in GR, the Schwarzschild spacetime can be formed by the collapse of a homogenous and isotropic dust perfect fluid 
\cite{Joshi}. We note that $\rho_H^{Im(0)} > 0$ for  
$$t_0 - \sqrt{\frac{2}{3G\pi \rho_H^{(0)}}} < t < t_0.$$

\subsection{Gravitational Collapse with $\lambda \not= 1$}

We now consider the case of $\lambda \neq 1$. For an exterior static spherically symmetric vacuum  spacetime with $\lambda \neq 1$ 
and $\Lambda_g = 0$, equation (\ref{3.3f}) implies that 
\bq
\nu^+ = -\frac{1}{2}\ln\biggl(1 - \frac{2B}{r} \biggr),
\eq
where $B$ is a constant. On the other hand, for the interior FLRW region, we have
\bq
\nu^- = -\frac{1}{2}\ln\biggl(1 - k\frac{r^2}{a^2(t)}\biggr).
\eq
Hence, the condition $\nu_{,t}^+ = \nu_{,t}^-$ on $\Sigma$ implies that
$0 = k\mathcal{R}(t)^2.$
Consequently, in order for a solution with $\mathcal{R}(t) \neq 0$ to exist, we must have $k = 0$. 
The conditions that $\nu$ and $\nu_{,r}$ be continuous across $\Sigma$ then reduce to
${2B}/{\mathcal{R}(t)}  = 0$.
Thus, in order for a nontrivial solution to exist we must have 
$k  = B = 0$.
Thus,  
we have
\bqn
\label{nunumu}
  \nu^- = \nu^+ = 0, \qquad \mu^- = \ln\big({- rH}\big).
\eqn  
On the other hand, since $\lambda \neq 1$, the momentum constraint (\ref{3.3c}) yields
\bqn
\lb{mu2}
   \mu^+(r) = \ln\biggl(C_1 r + \frac{C_2}{r^2}\biggr),
\eqn
where $C_1$ and $C_2$ are constants.
The field equations (\ref{3.3e}) - (\ref{3.3h}) are then satisfied provided that
\bqn
  A^+(r) = A_0^+ - \frac{3C_2^2}{8r^4} + \frac{3(1-3\lambda)C_1^2 + 2\Lambda}{8}r^2,
\eqn
where $A_0^+$ is a constant. It is interesting to note that this class of solutions was first found in \cite{LMW} in the IR limit. However,  
{since the restriction of the spacetime to the leaves $t = $ constant is flat, we have $R_{ij} = 0$, and the higher-order derivative terms {of $R_{ij}$ 
vanish identically, so} they are also solutions of the full theory.} 
Moreover, since
$$\mu_{,r}^+ = \frac{C_1 r^3 - 2C_2}{C_1 r^4 + C_2 r}, \qquad \mu_{,r}^- = \frac{1}{r},$$
we find
\bqn\label{hatmur2}
\hat{\mu}_{,r} = \frac{- 3C_2}{C_1 r^4 + C_2 r}.
\eqn
Thus, the requirement that $\mu$ is $C^1$ implies that $C_2 = 0$. The continuity of $\mu$ then requires that $H(t) = - C_1$ is a constant and so
$$a(t) = a_0 e^{-C_1 t}.$$
It follows that $\mu$ is smooth across $\Sigma$. 
Note also that
the asymptotical-flatness condition requires $C_1 = 0$. However, in the following we leave the possibility of $C_1 \neq 0$ open. 

We find that
\bqn \nb
&& \mathcal{L}_K^+ = 3C_1^2 (1-3 \lambda ),
\quad 
\mathcal{L}_V^+ = 2\Lambda, \quad \mathcal{L}_A^+ = 0,
	\\
&& v^+ = 0, \quad J_A^+ = 0, \quad J_\varphi^+ = 0, \quad \rho_H^+ = 0.	
\eqn
In order for the integral over the exterior region in the Hamiltonian constraint (\ref{distHamiltonianconstraint}) to converge, we also need to assume that
\bqn\label{C1Lambdaconstraint}
  3C_1^2(1 - 3\lambda) + 2\Lambda = 0.
\eqn
Thus, $A^+(r) = A_0^+$ is a constant and equation (\ref{aeq}) implies that $p^-(t) = 0$, that is, the perfect fluid in the interior region consists of dust.

Similar to the case with $\lambda = 1$,  the interior solution is still  of the form (\ref{flrwsolution2}), i.e.
\bqn\nb
&&J_A^- = J_\varphi^- = 0, \quad A^- = A^-(t).
\eqn
In view of (\ref{FLRWk0Ls}), we have
\bqn\nb
&&\mathcal{L}_\varphi^- = 0, \qquad \mathcal{L}_\lambda^- = 0, \qquad
\mathcal{L}_K^- = 3(1-3\lambda)H^2, 
	\\  
&& \mathcal{L}_V^- = 2\Lambda, \qquad \mathcal{L}_A^- = 0, \qquad v^- = 0.	
\eqn
We assume that the thin shell of matter separating the interior and exterior solutions is such that
\bqn\nb
&& p_r = 0, \quad v = 0, \quad J_\varphi = J_\varphi^{Im(0)} \delta(\Phi),
	\\\nb
&& p_\theta = p_\theta^{Im(0)} \delta(\Phi), \quad \rho_H = \rho_H^- + \rho_H^{Im(0)}\delta(\Phi), \qquad
	\\\label{shellassumptions2}
&& A = A^D + A^{Im(0)}\delta(\Phi), \quad J_A = 0, \quad \mathcal{L}_A = \mathcal{L}_A^-,
\eqn
with $\rho_H^- = \rho_H^-(t)$. 

\begin{proposition} 
For the spacetime defined by (\ref{nunumu})-(\ref{shellassumptions2}), the six junction conditions
 (\ref{distHamiltonianconstraint})-(\ref{distmomentumconservation}) reduce to the following six conditions:
\bqn  \label{Junc2Hamiltonianconstraint}
&& \rho_H^-(t) \frac{\mathcal{R}(t)}{3} + \rho_H^{Im(0)}(t, \mathcal{R}(t)) = 0,
	\\ \label{Junc2vDJDJD}
&& J_\varphi^{Im(0)} = 0,	
	\\ \lb{AAa}
&& \text{$A(t,r) = A_0$ is a constant},	
	\\ \label{Junc2dynamic2}
&&  p_\theta^{Im(0)} = 0 \text{  on $\Sigma$,} 
	\\ \nb
&& \frac{d}{dt}\Bigl[\rho_{H}^{Im(0)}(t, \mathcal{R}(t))\Bigr] +  2 \frac{\dot{\mathcal{R}}(t)}{\mathcal{R}(t)} \rho_H^{Im(0)} (t, \mathcal{R}(t))
	\\ \label{Junc2energyconservation}
&&   + \dot{\mathcal{R}}(t) \rho_H^-(t) + \frac{C_1 \rho_{H,t}^-(t) \mathcal{R}(t)^2}{4} = 0.
\eqn
\end{proposition}
\proofbegin
For the spacetime defined by (\ref{nunumu})-(\ref{shellassumptions2}), condition (\ref{distHamiltonianconstraint}) reduces to
\bqn\nb
&& \left( 3(1-3\lambda)C_1^2 + 2\Lambda + 4 \pi G \rho_H^-(t) \right) \int_0^{\mathcal{R}(t)}  r^2 dr
	\\  \nb
&& + \int_{\mathcal{R}(t)}^\infty 
\biggl(3C_1^2(1 - 3\lambda) + 2\Lambda\biggr) r^2 dr 
	\\ \nb
&& + 4\pi G \rho_H^{Im(0)} r^2 \Big|_{r = \mathcal{R}(t)}
= 0,
\eqn
which, in view of (\ref{C1Lambdaconstraint}), yields (\ref{Junc2Hamiltonianconstraint}). Moreover, equation (\ref{Jvarphidelta}) reduces to (\ref{Junc2vDJDJD}).

The functions $F$ and $F^A$ are given by (\ref{FrrFthetatheta}) - (\ref{FArrFAthetatheta}) also for $\lambda \neq 1$. 
Hence, condition (\ref{distdynamic1}) implies that $A$ is continuous across $\Sigma$ just like in the case of $\lambda = 1$. Since $A^+ = A_0^+$ is constant and $A^-(t)$ is independent of $r$, this gives (\ref{AAa}).
Condition (\ref{distdynamic2}) then reduces to
\bqn \nb
&& \frac{1}{r} F_{\theta\theta}^{AIm(0)} \delta(\Phi) 
+ \frac{1}{r} F_{\theta\theta}^{AIm(1)} \delta'(\Phi) 
	\\ \nb
&& + \frac{1}{r} F_{\theta\theta}^{AIm(2)}  \delta''(\Phi) + 8\pi G r p_\theta^{Im(0)} \delta(\Phi) = 0.
\eqn
Since $A$ is a constant, this yields (\ref{Junc2dynamic2}).

Conditions (\ref{distenergyconservation}) and (\ref{distmomentumconservation}) reduce to (\ref{Junc2energyconservation}) and (\ref{Junc2dynamic2}).
\proofend
   
Conditions (\ref{Junc2Hamiltonianconstraint}) and (\ref{Junc2energyconservation}) imply that
$$\dot{\rho}_{H}^-(t) \mathcal{R}(t) \bigg(\frac{C_1}{4}\mathcal{R}(t) - \frac{1}{3}\bigg) = 0.$$
Excluding the case of no collapse where $\mathcal{R}(t)$ is a constant, it follows that $\rho_H^-$ must be a constant.

In summary, in the case $\lambda \not= 1$ a static spherical spacetime can be produced by gravitational collapse of a  homogeneous and isotropic
dust fluid. However, the space-time outside of such a fluid is not asymptotically flat, as one can see from
Eqs.(\ref{nunumu}) and (\ref{mu2}) with $C_2 = 0$.

\section{Conclusions}
\nequation

In this paper, we have studied gravitational collapse of  a spherical cloud of fluid  with a finite radius  in the framework of the nonrelativistic 
general covariant theory of the HL gravity with the projectability condition and an arbitrary coupling constant $\lambda$.  
Using distribution theory, we have developed the general junction conditions for such a collapsing spherical body, with the minimal 
requirement that {\em such junctions should be mathematically meaningful in the sense  of generalized functions.}
The general junction conditions have been  summarized in Table I.  

As one of the simplest applications, we have studied  a collapsing star that is made of a homogeneous and isotropic perfect fluid, while the external  {region} is described 
by a stationary  spacetime. We have found that the problem reduces to the matching of six independent conditions that the Arnowitt-Deser-Misner  
variables ($N, N^i, g_{ij}$) and the gauge field $A$ and Newtonian prepotential $\varphi$ must satisfy. 

For the case of  a homogeneous and isotropic dust fluid (a perfect fluid with vanishing pressure), we have found explicitly the space-time outside of the collapsing
sphere. In particular,   in the case $\lambda  = 1$,  the external spacetimes are described by the Schwarzschild (anti-) de Sitter solutions, written in the Painlev\'e-Gullstrand
coordinates \cite{GP}.  It is remarkable that the collapse of a homogeneous and isotropic dust to a 
 Schwarzschild black hole, studied by Oppenheimer and Snyder in general relativity more than 80 years ago \cite{OppSnyder}, is a particular case. However, there are 
 fundamental differences. First, in general relativity a thin shell does not necessarily appear on the surface of the collapsing sphere \cite{OppSnyder}, while in the current case we have shown that
 such a thin shell must exist, as one can see from Propositions V.2 - 4 given in Section V. Second, in general relativity because of the local conservation of  energy of the collapsing boy, the energy density of the dust fluid is inversely proportional to
 the cube of the radius of the fluid, while in the current case it remains a constant, as now the conservation law becomes a global one [cf. Eq.(\ref{eq1})], and the energy of the collapsing star is not 
 necessarily  conserved locally.
 
  In the case $\lambda \not= 1$, the space-time outside of the homogeneous and isotropic dust fluid is described by Eqs.(\ref{nunumu}) and (\ref{mu2}) with $C_2 = 0$. 
 It is clear that such a space-time is not asymptotically flat. Therefore, in this case to obtain an asymptotically flat space-time outside of a collapsing dust fluid,  it must not be
 homogeneous and/or isotropic. 
 
 From the above simple examples, one can already see the significant differences between the HL theory and general relativity in the strong gravitational field regime. Therefore, it is very interesting to study 
 gravitational collapse of more general fluids, such as perfect fluids with different equations of state, anisotropic fluids with or without heat flows. Particular attentions should be paid on the roles that the equation of state and heat flows might play. It would be extremely interesting to study  the implications  to black hole physics, or more general to (observational) astrophysics and cosmology \cite{Joshi}. Since the general formulas have been already laid down in this paper, we expect that such studies  can be carried easily.

 As emphasized previously,  our treatments    for the junction conditions of a collapsing star presented in this paper can be easily generalized to
  other models of the Ho\v{r}ava-Lifshitz gravity,  or more general to any model of high-order derivative gravity theories.

   
~\\{\bf Acknowledgments:}  We would like to express our gratitude to Jie Yang for his valuable comments and discussions. 
 The work of AW was supported in part by DOE  Grant, DE-FG02-10ER41692. 
JL acknowledges support from the EPSRC, UK.

\section*{Appendix A:  Functions $\left(F_{s}\right)_{ij}$ and $F_{(\varphi, n)}^{ij}$ }
\renewcommand{\theequation}{A.\arabic{equation}} \setcounter{equation}{0}

The geometric 3-tensors  $F^{ij}$ and $F_{(\varphi, n)}^{ij}$ defined in  Eq.(\ref{eq3a})
are given  by
  \bqn \lb{A.1}
(F_{0})_{ij}&=& -\frac12g_{ij},\nb\\
(F_{1})_{ij}&=&-\frac12g_{ij}R+R_{ij},\nb\\
(F_{2})_{ij} &=&-\frac12g_{ij}R^2+2RR_{ij}-2\nabla_{(i}\nabla_{j)}R\nb\\
                     & & +2g_{ij}\nabla^2R,\nb\\
(F_{3})_{ij}&=&-\frac12g_{ij}R_{mn}R^{mn}+2R_{ik}R^k_j-2\nabla^k\nabla_{(i}R_{j)k}\nb\\
                     && +\nabla^2R_{ij}+g_{ij}\nabla_m\nabla_nR^{mn},\nb\\
(F_{4})_{ij}&=&-\frac12g_{ij}R^3+3R^2R_{ij}-3\nabla_{(i}\nabla_{j)}R^2\nb\\
                    && +3g_{ij}\nabla^2R^2,\nb\\
(F_{5})_{ij}&=&-\frac12g_{ij}RR^{mn}R_{mn}+R_{ij}R^{mn}R_{mn}\nb\\
                    &&+ 2RR_{ki}R^k_j  -\nabla_{(i}\nabla_{j)}\left(R^{mn}R_{mn}\right)\nb\\
                    && - 2\nabla^n\nabla_{(i}RR_{j)n}  +g_{ij}\nabla^2\left(R^{mn}R_{mn}\right)\nb\\
                    &&  + \nabla^2\left(RR_{ij}\right)   +g_{ij}\nabla_m\nabla_n\left(RR^{mn}\right),\nb\\
(F_{6})_{ij}&=&-\frac12g_{ij}R^m_nR^n_pR^p_m+3R^{mn}R_{ni}R_{mj}\nb\\
                      && +\frac32\nabla^2\left(R_{in}R^n_j\right)   + \frac32g_{ij}\nabla_k\nabla_l\left(R^k_nR^{ln}\right)\nb\\
                      & & -3\nabla_k\nabla_{(i}\left(R_{j)n}R^{nk}\right),\nb\\
(F_{7})_{ij}&=&\frac12g_{ij}(\nabla R)^2- \left(\nabla_iR\right)\left(\nabla_jR\right) + 2R_{ij}\nabla^2R\nb\\
&& -2\nabla_{(i}\nabla_{j)}\nabla^2R+2g_{ij}\nabla^4R,\nb\\
(F_{8})_{ij}&=& -\frac12g_{ij}\left(\nabla_p R_{mn}\right)\left(\nabla^p R^{mn}\right) -\nabla^4R_{ij}\nb\\
&&  + \left(\nabla_i R_{mn}\right)\left(\nabla_j R^{mn}\right) +2\left(\nabla_p R_{in}\right)\left(\nabla^p R^n_j\right)\nb\\
&&  +2\nabla^n\nabla_{(i}\nabla^2R_{j)n}+2\nabla_n\left(R^n_m\nabla_{(i}R^m_{j)}\right)\nb\\
&& -2\nabla_n\left(R_{m(j}\nabla_{i)}R^{mn}\right)-2\nabla_n\left(R_{m(i}\nabla^nR^m_{j)}\right)\nb\\
 && -g_{ij}\nabla^n\nabla^m\nabla^2R_{mn},\\
  \lb{A.2}
F_{(\varphi, 1)}^{ij} &=& \frac{1}{2}\varphi\left\{\Big(2K + \nabla^{2}\varphi\Big) R^{ij}  
- 2 \Big(2K^{j}_{k} + \nabla^{j} \nabla_{k}\varphi\Big) R^{ik} \right.\nb\\
& & ~~~~~ - 2 \Big(2K^{i}_{k} + \nabla^{i} \nabla_{k}\varphi\Big) R^{jk}\nb\\
& &~~~~~\left. 
- \Big(2\Lambda_{g} - R\Big) \Big(2K^{ij} + \nabla^{i} \nabla^{j}\varphi\Big)\right\},\nb\\
F_{(\varphi, 2)}^{ij} &=& \frac{1}{2}\nabla_{k}\left\{\varphi{\cal{G}}^{ik}  
\Big(\frac{2N^{j}}{N} + \nabla^{j}\varphi\Big) \right. \nb\\
& & \left.
+ \varphi{\cal{G}}^{jk}  \Big(\frac{2N^{i}}{N} + \nabla^{i}\varphi\Big) 
-  \varphi{\cal{G}}^{ij}  \Big(\frac{2N^{k}}{N} + \nabla^{k}\varphi\Big)\right\}, \nb\\   
F_{(\varphi, 3)}^{ij} &=& \frac{1}{2}\left\{2\nabla_{k} \nabla^{(i}f^{j) k}_{\varphi}  
- \nabla^{2}f_{\varphi}^{ij}   - \left(\nabla_{k}\nabla_{l}f^{kl}_{\varphi}\right)g^{ij}\right\},\nb\\
\eqn
where
\bqn
\lb{A.3}
f_{\varphi}^{ij} &=& \varphi\left\{\Big(2K^{ij} + \nabla^{i}\nabla^{j}\varphi\Big) 
- \frac{1}{2} \Big(2K + \nabla^{2}\varphi\Big)g^{ij}\right\}.\nb\\
\eqn

{The $F_{ij}$ for the spherical spacetime (\ref{3.1b}) are found to be}
%
\bqn
\lb{A.4}
(F_0)_{ij} & = & -\frac{e^{2 \nu }}{2} \delta_i^r\delta_j^r
-\frac{r^2}{2} \Omega_{ij},\nb\\
(F_1)_{ij} & = & \frac{1-e^{2 \nu }}{r^2} \delta_i^r\delta_j^r
-e^{-2 \nu } r \nu ' \Omega_{ij},\nb\\
(F_2)_{ij} & = & -\frac{2 e^{-2 \nu } }{r^4} \Big[6 e^{2 \nu }+e^{4 \nu }-8 r^2 \nu ''\nb\\
&&  ~~~~~~~~~~~~~ +12 r^2  \left(\nu '\right)^2-7\Big] \delta_i^r\delta_j^r
   	\nb\\
&& +\frac{2 e^{-4 \nu }}{r^2} \Big[6 e^{2 \nu }+e^{4 \nu }+4 \nu ^{(3)} r^3+24
   r^3 \left(\nu '\right)^3\nb\\
   &&  -2 r \nu ' \left(-3 e^{2 \nu }+14 r^2 \nu
   ''+7\right)-7\Big]\Omega_{ij},
	\nb\\
(F_3)_{ij} & = & -\frac{e^{-2 \nu }}{r^4} \Big[4 e^{2 \nu }+e^{4 \nu }-6 r^2 \nu ''\nb\\
&& ~~~~~~~~~~~ +9 r^2
   \left(\nu '\right)^2-5\Big]  \delta_i^r\delta_j^r
   	\nb\\
&& + \frac{e^{-4 \nu }}{r^2} \Big[4 e^{2 \nu }+e^{4 \nu }+3 \nu ^{(3)} r^3+18 r^3
   \left(\nu '\right)^3\nb\\
   && -r \nu ' \left(-4 e^{2 \nu }+21 r^2 \nu
   ''+10\right)-5\Big] \Omega_{ij},
	\nb\\
(F_4)_{ij} & = & -\frac{4 e^{-4 \nu } }{r^6} \left(e^{2 \nu }+2 r \nu '-1\right) \Big[22 e^{2
   \nu }+e^{4 \nu }-24 r^2 \nu ''\nb\\
   && +40 r^2 \left(\nu '\right)^2-2
   \left(e^{2 \nu }-1\right) r \nu '-23\Big] \delta_i^r\delta_j^r
   	\nb\\
&& + \frac{4 e^{-6 \nu } }{r^4} \biggl\{ 240 r^4 \left(\nu '\right)^4+4 \left(18 e^{2
   \nu }-17\right) r^3 \left(\nu '\right)^3\nb\\
   && -12 r^2 \left(\nu '\right)^2
   \left(-11 e^{2 \nu }+22 r^2 \nu ''+15\right)
   	\nb\\
&&  +3 r \nu ' \Big[-6 e^{2
   \nu }+7 e^{4 \nu }+8 \nu ^{(3)} r^3\nb\\
   && ~~~~~~~~~~ -28 \left(e^{2 \nu }-1\right) r^2  \nu ''-1\Big]
   	\nb\\
&&    +2 \Big[12 r^4 \left(\nu ''\right)^2-24 \left(e^{2 \nu }-1\right) r^2 \nu''\nb\\
&& +\left(e^{2 \nu
   }-1\right) \left(22 e^{2 \nu }+e^{4 \nu }+6 \nu ^{(3)}
   r^3-23\right)\Big]\biggr\} \Omega_{ij},
	\nb\\
(F_5)_{ij} & = & -\frac{2 e^{-4 \nu }}{r^6}\biggl\{60 r^3 \left(\nu '\right)^3\nb\\
&& +\left(e^{2 \nu
   }-1\right) \left(16 e^{2 \nu }+e^{4 \nu }-14 r^2 \nu
   ''-17\right)\nb\\
   && +\left(21 e^{2 \nu }-17\right) r^2 \left(\nu
   '\right)^2
   	\nb\\
&&  -4 r \nu ' \left(-7 e^{2 \nu }+9 r^2 \nu
   ''+7\right)\biggr\} \delta_i^r\delta_j^r
   	\nb\\
&& + \frac{2 e^{-6 \nu }}{r^4} \biggl\{18 r^4 \left(\nu ''\right)^2+180 r^4
   \left(\nu '\right)^4\nb\\
   && +\left(e^{2 \nu }-1\right) \left(32 e^{2 \nu }+2
   e^{4 \nu }+7 \nu ^{(3)} r^3-34\right)
   	\nb\\
&&   +21 \left(2 e^{2 \nu }-1\right)
   r^3 \left(\nu '\right)^3-28 \left(e^{2 \nu }-1\right) r^2 \nu ''\nb\\
   && -r^2
   \left(\nu '\right)^2 \left(-77 e^{2 \nu }+198 r^2 \nu
   ''+101\right)
   	\nb\\
&&   +r \nu ' \Big[3 \left(-8 e^{2 \nu }+5 e^{4 \nu }+6 \nu
   ^{(3)} r^3+3\right)\nb\\
   && ~~~~~~~~ -\left(49 e^{2 \nu }-41\right) r^2 \nu
   ''\Big]\biggr\} \Omega_{ij},
	\nb\\
(F_6)_{ij} & = & \frac{e^{-4 \nu } }{r^6} \biggl\{ -50 r^3 \left(\nu '\right)^3\nb\\
&& -\left(e^{2 \nu
   }-1\right) \left(13 e^{2 \nu }+e^{4 \nu }-6 r^2 \nu ''-14\right)\nb\\
   && -9
   e^{2 \nu } r^2 \left(\nu '\right)^2
	 +6 r \nu ' \left(-2 e^{2 \nu }+5
   r^2 \nu ''+2\right)\biggr\} \delta_i^r\delta_j^r
   	\nb\\
&& + \frac{e^{-6 \nu }}{r^4}\biggl\{15 r^4 \left(\nu ''\right)^2+150 r^4 \left(\nu
   '\right)^4\nb\\
   && +\left(e^{2 \nu }-1\right) \left(26 e^{2 \nu }+2 e^{4 \nu
   }+3 \nu ^{(3)} r^3-28\right)
   	\nb\\
&&    +\left(18 e^{2 \nu }+25\right) r^3
   \left(\nu '\right)^3-12 \left(e^{2 \nu }-1\right) r^2 \nu ''\nb\\
   && -3 r^2
   \left(\nu '\right)^2 \left(-11 e^{2 \nu }+55 r^2 \nu ''+12\right)
   	\nb\\
&&   +3
   r \nu ' \Big[-12 e^{2 \nu }+4 e^{4 \nu }+5 \nu ^{(3)} r^3\nb\\
   &&~~~~~~~~~~ -\left(7
   e^{2 \nu }-1\right) r^2 \nu ''+8\Big]\biggr\} \Omega_{ij},
	\nb\\
(F_7)_{ij} & = & \frac{8 e^{-4 \nu }}{r^6} \biggl\{6 e^{2 \nu }+e^{4 \nu }-2 \nu ^{(4)} r^4+15
   r^4 \left(\nu ''\right)^2\nb\\
   && +40 r^4 \left(\nu '\right)^4 +4 r \nu '
   \left(2 e^{2 \nu }+5 \nu ^{(3)} r^3-6\right)
   	\nb\\
&&   -2 r^2 \left(\nu '\right)^2 \left(-3 e^{2 \nu
   }+41 r^2 \nu ''+15\right)\nb\\
   &&  -4 \left(e^{2 \nu
   }-3\right) r^2 \nu '' -7 \biggr\} \delta_i^r\delta_j^r
   	\nb\\
&&-\frac{8 e^{-6 \nu }}{r^4} \biggl\{12 e^{2 \nu }+2 e^{4 \nu }+\nu ^{(5)}
   r^5\nb\\
   && +120 r^5 \left(\nu '\right)^5+2 e^{2 \nu } \nu ^{(3)} r^3-6 \nu
   ^{(3)} r^3
   	\nb\\
&&   +r \nu ' \Big[24 e^{2 \nu }+e^{4 \nu }-16 \nu ^{(4)}
   r^4+127 r^4 \left(\nu ''\right)^2\nb\\
   && -2 \left(7 e^{2 \nu }-33\right) r^2
   \nu ''-57\Big]
   	\nb\\
&&  -r^2 \nu '' \left(8 e^{2 \nu }+25 \nu ^{(3)}
   r^3-24\right)\nb\\
   && +r^2 \left(\nu '\right)^2 \left(22 e^{2 \nu }+101 \nu
   ^{(3)} r^3-102\right)
   	\nb\\
&&    -2 r^3 \left(\nu '\right)^3 \left(-6 e^{2 \nu
   }+163 r^2 \nu ''+45\right)-14\biggr\}\Omega_{ij},
	\nb\\
(F_8)_{ij} & = &\frac{e^{-4 \nu }}{r^6} \biggl\{6 \left(2 e^{2 \nu }+e^{4 \nu }-\nu ^{(4)}
   r^4-3\right)\nb\\
   && +45 r^4 \left(\nu ''\right)^2+120 r^4 \left(\nu
   '\right)^4+10 r^3 \left(\nu '\right)^3
   	\nb\\
&&    -8 \left(e^{2 \nu }-4\right)
   r^2 \nu ''\nb\\
   && -r^2 \left(\nu '\right)^2 \left(-12 e^{2 \nu }+246 r^2 \nu
   ''+77\right)
   	\nb\\
&&    +2 r \nu ' \left(8 e^{2 \nu }+30 \nu ^{(3)} r^3-3 r^2
   \nu ''-32\right)\biggr\} \delta_i^r\delta_j^r
   	\nb\\
&& + \frac{e^{-6 \nu }}{r^4} \biggl\{-24 e^{2 \nu }-12 e^{4 \nu }-3 \nu ^{(5)}
   r^5\nb\\
   && -360 r^5 \left(\nu '\right)^5-3 r^4 \left(\nu ''\right)^2-30 r^4
   \left(\nu '\right)^4
   	\nb\\
  &&  -4 e^{2 \nu } \nu ^{(3)} r^3+16 \nu ^{(3)}
   r^3\nb\\
   && +r^2 \nu '' \left(16 e^{2 \nu }+75 \nu ^{(3)} r^3-64\right)
   	\nb\\
&&    +2 r^3
   \left(\nu '\right)^3 \left(-12 e^{2 \nu }+489 r^2 \nu
   ''+113\right)\nb\\
   && -r^2 \left(\nu '\right)^2 \left(44 e^{2 \nu }+303 \nu
   ^{(3)} r^3-33 r^2 \nu ''-269\right)
   	\nb\\
&&    -r \nu ' \Big[381 r^4 \left(\nu
   ''\right)^2-2 \left(14 e^{2 \nu }-85\right) r^2 \nu ''\nb\\
   && +6
   \left(8 e^{2 \nu }+e^{4 \nu }-8 \nu ^{(4)} r^4-25\right)\nb\\
   && +3\nu ^{(3)}
   r^3\Big]+36\biggr\} \Omega_{ij}.
\eqn
where $\nu' = \partial \nu/\partial r$ and $\Omega_{ij} = \delta_i^\theta \delta_j^\theta + \sin^2\theta \delta_i^\phi \delta_j^\phi$.

\section*{Appendix B: Proof of (\ref{Fpm0})-(\ref{Fdeltaeqs})}
\renewcommand{\theequation}{B.\arabic{equation}} \setcounter{equation}{0}
Let $F$ be given by (\ref{Fdef}). We will show that the equation $F = 0$ is equivalent to the conditions (\ref{Fpm0}) and (\ref{Fdeltaeqs}).
It is clear that the equation $F= 0$ is equivalent to (\ref{Fpm0}) together with the condition
\bqn\label{sumdelta0}
  \sum_{k = 0}^n F^{Im(k)} \delta^{(k)}(\Phi) = 0.
\eqn  
It remains to show that (\ref{sumdelta0}) is equivalent to (\ref{Fdeltaeqs}). 

Suppose first that (\ref{sumdelta0}) holds. Then, multiplying (\ref{sumdelta0}) by $\Phi^{n-j}$ and using the recursion relation (\ref{deltarecursion}) repeatedly, we find
\bq\label{GdeltaPhi}
  G \delta^{(j)}(\Phi) = 0, \qquad 0 \leq j \leq n,
\eq  
where the function $G$ is defined in a neighborhood of $\Sigma$ by
\bq\label{Gdef}
G(x) = \sum_{k = 0}^n (-1)^{k} k! F^{Im(k)}(x) \Phi^{n-k}(x).
\eq
Equation (\ref{GdeltaPhi}) with $j = 0$ implies that the restriction of $G$ to $\Sigma$ vanishes, i.e. $G|_\Sigma = 0$.
Equation (\ref{GdeltaPhi}) with $j = 1$ then gives
$$0 = G \delta'(\Phi) = \frac{G}{\Phi} \Phi \delta'(\Phi)
= -\frac{G}{\Phi} \delta(\Phi) \quad \text{i.e.} \quad \frac{G}{\Phi}\Big|_\Sigma = 0.$$
In terms of local coordinates $\{u^j\}$ such that $u^1 = \Phi$ while the remaining coordinates $\{u^j\}_{j \geq 2}$ parametrize the level surfaces of $\Phi$, we have
$$0 = \frac{G}{\Phi}\Big|_\Sigma = \frac{\partial G}{\partial \Phi}\bigg|_{\Phi = 0}.$$
Thus, $G$ vanishes to the first order on� $\Sigma$. Repeating the above procedure $n$ times, we infer that $G$ vanishes to the $n$th order on $\Sigma$:
\bq\label{partialjGzero}
\frac{G}{\Phi^j}\bigg|_{\Sigma} = 0 \quad \text{i.e.}\quad 
\frac{\partial^j G}{\partial \Phi^j}\bigg|_{\Phi = 0} = 0, \qquad 0 \leq j \leq n.
\eq
The partial derivatives denoted in the local coordinates $(u^j)$ by $\frac{\partial^j}{\partial \Phi^j}$ can be expressed invariantly as in (\ref{partialPhij}). 
Substituting the expression (\ref{Gdef}) for $G$ into (\ref{partialjGzero}), we find
\bqn\nb
0 &=& \sum_{k = 0}^n (-1)^{k} k! \sum_{r=0}^j {j \choose r} \frac{\partial^{j-r} F^{Im(k)}}{\partial\Phi^{j-r}} \frac{\partial^{r}\Phi^{n-k}}{\partial \Phi^r}\bigg|_{\Phi = 0}
	\\ \nb
&=& \sum_{k = n-j}^n (-1)^{k} k! (n-k)! {j \choose n-k} \frac{\partial^{j-(n-k)} F^{Im(k)}}{\partial\Phi^{j-(n-k)}}\bigg|_{\Phi = 0},
	\\ \nb
&&\hspace{4cm} \qquad 0 \leq j \leq n.
\eqn	
Replacing $k$ by $n-k$, we find (\ref{Fdeltaeqs}).

Conversely, if (\ref{Fdeltaeqs}) holds, then tracing the above steps backwards, we infer that (\ref{GdeltaPhi}), and hence also (\ref{sumdelta0}), holds.

\end{document}